\documentclass[twocolumn]{aastex701}

\usepackage{booktabs}
\usepackage{amsmath,amssymb}
\usepackage{comment}
\usepackage{hyperref}

\begin{document}
\suppressAffiliations
\title{Large-Scale Structure in COSMOS-Web: Tracing Galaxy Evolution \\in the Cosmic Web up to $z \sim 7$ with the Largest JWST Survey}

\shorttitle{Large-Scale Structure in COSMOS-Web}
\shortauthors{Hatamnia et al.}

\correspondingauthor{Hossein Hatamnia}
\email{hossein.hatamnia@email.ucr.edu}

\author[0009-0007-3673-4523]{Hossein Hatamnia}
\affiliation{Department of Physics and Astronomy, University of California, Riverside, 900 University Avenue, Riverside, CA 92521, USA}
\email{hossein.hatamnia@email.ucr.edu}

\author[0000-0001-5846-4404]{Bahram Mobasher}
\affiliation{Department of Physics and Astronomy, University of California, Riverside, 900 University Avenue, Riverside, CA 92521, USA}
\email{bahram.mobasher@ucr.edu}

\author[orcid=0000-0003-0749-4667]{Sina Taamoli}
\affiliation{Department of Physics and Astronomy, University of California, Riverside, 900 University Avenue, Riverside, CA 92521, USA}
\email{sina.taamoli@email.ucr.edu}

\author[0000-0001-9187-3605]{Jeyhan S. Kartaltepe}
\email{jeyhan@astro.rit.edu}
\affiliation{Laboratory for Multiwavelength Astrophysics, School of Physics and Astronomy, Rochester Institute of Technology, 84 Lomb Memorial Drive, Rochester, NY 14623, USA}

\author[0000-0002-0930-6466]{Caitlin M. Casey}
\email{cmcasey@ucsb.edu}
\affiliation{Department of Physics, University of California, Santa Barbara, Santa Barbara, CA 93106, USA}
\affiliation{Cosmic Dawn Center (DAWN), Denmark}

\author[0000-0003-3596-8794]{Hollis B. Akins}
\email{hollis.akins@gmail.com}
\affiliation{Department of Astronomy, The University of Texas at Austin, 2515
Speedway Blvd Stop C1400, Austin, TX 78712, USA}
\email{hollis.akins@utexas.edu}

\author[0000-0002-0245-6365]{Malte Brinch}
\email{malte.brinch@uv.cl}
\affiliation{Instituto de Física y Astronomía, Universidad de Valparaíso, Avda. Gran Bretan\~{a} 1111, Valparaíso, Chile}
\affiliation{Millennium Nucleus for Galaxies (MINGAL)}

\author[0000-0003-3691-937X]{Nima Chartab}
\email{nima.chartab@email.ucr.edu}
\affiliation{Caltech/IPAC, MS 314-6, 1200 E. California Blvd. Pasadena, CA 91125, USA}

\author[0000-0003-4761-2197]{Nicole E. Drakos}
\email{ndrakos@hawaii.edu}
\affiliation{Department of Physics and Astronomy, University of Hawaii, Hilo, 200 W Kawili St, Hilo, HI 96720, USA}

\author[0000-0002-9382-9832]{Andreas L. Faisst}
\email{afaisst@caltech.edu}
\affiliation{Caltech/IPAC, MS 314-6, 1200 E. California Blvd. Pasadena, CA 91125, USA}

\author[0000-0001-8519-1130]{Steven L. Finkelstein}
\affiliation{Department of Astronomy, The University of Texas at Austin, 2515
Speedway Blvd Stop C1400, Austin, TX 78712, USA}
\affiliation{Cosmic Frontier Center, The University of Texas at Austin, Austin, TX, USA}
\email{stevenf@astro.as.utexas.edu}

\author[0000-0002-3560-8599]{Maximilien Franco}
\email{maximilien.franco@cea.fr}
\affiliation{Université Paris-Saclay, Université Paris Cité, CEA, CNRS, AIM, 91191 Gif-sur-Yvette, France}

\author[0009-0003-2158-1246]{Finn Giddings}
\email{finndg@hawaii.edu}
\affiliation{Institute for Astronomy, University of Hawai‘i, 2680 Woodlawn Drive, Honolulu, HI 96822, USA}

\author[0000-0002-0236-919X]{Ghassem Gozaliasl}
\email{ghassem.gozaliasl@gmail.com}
\affiliation{Department of Computer Science, Aalto University, P.O. Box 15400, FI-00076 Espoo, Finland}
\affiliation{Department of Physics, University of Helsinki, P.O. Box 64, FI-00014 Helsinki, Finland}

\author[0009-0003-3097-6733]{Ali Hadi}
\email{ahadi005@ucr.edu}
\affiliation{Department of Physics and Astronomy, University of California, Riverside, 900 University Avenue, Riverside, CA 92521, USA}

\author[0009-0006-3071-7143]{Aryana Haghjoo}
\email{ahagh010@ucr.edu}
\affiliation{Department of Physics and Astronomy, University of California, Riverside, 900 University Avenue, Riverside, CA 92521, USA}

\author[0000-0003-0129-2079]{Santosh Harish}
\email{harish.santosh@gmail.com}
\affiliation{Laboratory for Multiwavelength Astrophysics, School of Physics and Astronomy, Rochester Institute of Technology, 84 Lomb Memorial Drive, Rochester, NY 14623, USA}

\author[0000-0002-7303-4397]{Olivier Ilbert}
\email{olivier.ilbert@lam.fr}
\affiliation{Aix Marseille Univ, CNRS, CNES, LAM, Marseille, France}

\author[0000-0002-9655-1063]{Pascale L. Jablonka}
\email{}
\affiliation{Ecole Polytechnique F\'ed\'erale de Lausanne, CH-1015 Lausanne, Switzerland}

\author[0000-0002-8412-7951]{Shuowen Jin}
\email{shuowen.jin@gmail.com}
\affiliation{Cosmic Dawn Center (DAWN), Denmark}
\affiliation{DTU Space, Technical University of Denmark, Elektrovej 327, 2800 Kgs. Lyngby, Denmark}

\author[0000-0002-0101-336X]{Ali Ahmad Khostovan}
\email{akhostov@gmail.com}
\affiliation{Department of Physics and Astronomy, University of Kentucky, 505 Rose Street, Lexington, KY 40506, USA}
\affiliation{Laboratory for Multiwavelength Astrophysics, School of Physics and Astronomy, Rochester Institute of Technology, 84 Lomb Memorial Drive, Rochester, NY 14623, USA}

\author[0000-0002-6610-2048]{Anton M. Koekemoer}
\email{koekemoer@stsci.edu}
\affiliation{Space Telescope Science Institute, 3700 San Martin Drive,
Baltimore, MD 21218, USA}

\author[0000-0002-0322-6131]{Ronaldo Laishram}
\email{}
\affiliation{National Astronomical Observatory of Japan, 2-21-1 Osawa, Mitaka, Tokyo 181-8588, Japan}

\author[0000-0001-9773-7479]{Daizhong Liu}
\email{dzliu@pmo.ac.cn}
\affiliation{Purple Mountain Observatory, Chinese Academy of Sciences, 10 Yuanhua Road, Nanjing 210023, China}

\author[0000-0002-3517-2422]{Matteo Maturi} 
\email{maturi@uni-heidelberg.de}
\affiliation{Zentrum f\"{u}r Astronomie, Universit\"{a}t Heidelberg, Philosophenweg 12, D-69120, Heidelberg, Germany}
\affiliation{Institute for Theoretical Physics, Philosophenweg 16, D-69120 Heidelberg, Germany}

\author[0000-0002-9489-7765]{Henry Joy McCracken}
\email{hjmcc@iap.fr}
\affiliation{Institut d’Astrophysique de Paris, UMR 7095, CNRS, and Sorbonne Université, 98 bis boulevard Arago, F-75014 Paris, France}

\author[0000-0001-9189-7818]{Crystal L. Martin}
\email{}
\affiliation{Department of Physics, University of California, Santa Barbara, Santa Barbara, CA 93106, USA}

\author[0000-0002-3473-6716]{Lauro Moscardini}
\email{lauro.moscardini@unibo.it}
\affiliation{University of Bologna - Department of Physics and Astronomy “Augusto Righi” (DIFA), Via Gobetti 93/2, I-40129 Bologna, Italy}
\affiliation{INAF- Osservatorio di Astrofisica e Scienza dello Spazio, Via Gobetti 93/3, I-40129, Bologna, Italy}
\affiliation{INFN- Sezione di Bologna, Viale Berti Pichat 6/2, I-40127, Bologna, Italy}

\author[0000-0001-8450-7885]{Diana Scognamiglio}
\affiliation{Jet Propulsion Laboratory, California Institute of Technology, 4800 Oak Grove Drive, Pasadena, CA 91109, USA}
\email{diana.scognamiglio@jpl.nasa.gov}

\author[0000-0002-7087-0701]{Marko Shuntov}
\email{marko.ov@nbi.ku.dk}
\affiliation{Cosmic Dawn Center (DAWN), Denmark}

\author[0009-0005-3133-1157]{Greta Toni}
\email{greta.toni4@unibo.it}
\affiliation{University of Bologna - Department of Physics and Astronomy “Augusto Righi” (DIFA), Via Gobetti 93/2, I-40129 Bologna, Italy}
\affiliation{INAF- Osservatorio di Astrofisica e Scienza dello Spazio, Via Gobetti 93/3, I-40129, Bologna, Italy}
\affiliation{Zentrum f\"{u}r Astronomie, Universit\"{a}t Heidelberg, Philosophenweg 12, D-69120, Heidelberg, Germany}

\author[0000-0002-6219-5558]{Alexander de la Vega}
\affiliation{Department of Physics and Astronomy, University of California, Riverside, 900 University Avenue, Riverside, CA 92521, USA}
\email{alexandd@ucr.edu}

\author[0000-0003-1614-196X]{John R. Weaver}
\thanks{Brinson Prize Fellow}
\affiliation{MIT Kavli Institute for Astrophysics and Space Research, 70 Vassar Street, Cambridge, MA 02139, USA}
\email{john.weaver.astro@gmail.com}

\author[0000-0002-8434-880X]{Lilan Yang}
\email{lxysps@rit.edu}
\affiliation{Laboratory for Multiwavelength Astrophysics, School of Physics and Astronomy, Rochester Institute of Technology, 84 Lomb Memorial Drive, Rochester, NY 14623, USA}

\begin{abstract}

We present a reconstruction of the large-scale structure using the James Webb Space Telescope’s (JWST) COSMOS-Web program to trace environmentally driven galaxy evolution up to $z\sim7$. We applied a weighted kernel density estimation method to 160,000 galaxies with robust photometric redshifts. We find that stellar mass has a positive correlation with density at all redshifts, stronger for quiescent galaxies (QGs) at $z\lesssim2.5$, while at higher redshifts ($2.5\lesssim z\lesssim5.5$) this trend is confined to extreme overdense environments, consistent with early mass assembly in proto-clusters. The star-formation rate (SFR) shows a negative trend with density for QGs at $z\lesssim1.2$, reversing at $z\gtrsim1.8$, while star-forming galaxies (SFGs) show a mild positive correlation up to $z\sim5.5$. The specific SFR remains nearly flat for SFGs and declines with density for QGs at $z\lesssim1.2$. Moreover, mass and environmental quenching efficiencies show that mass-driven processes dominate at $z\gtrsim2.5$, the two processes act with comparable strength between $0.8\lesssim z\lesssim2.5$, and environmental quenching becomes stronger for low-mass galaxies ($M_\star\lesssim10^{10}M_\odot$) at $z\lesssim0.8$. These findings reveal that large-scale structure drives galaxy evolution by enhancing early mass assembly in dense regions and increasingly suppressing star formation in low-mass systems at later times, establishing the environmental role of the cosmic web across cosmic history. COSMOS-Web, the largest JWST survey, provides accurate and deep photometric redshifts, reaching $80\%$ mass completeness at $\log{M_\star/M_\odot}\sim8.7$ at $z\sim7$, enabling the first view of how environments shaped galaxy evolution from the epoch of reionization to the present day.

\end{abstract}


\section{Introduction} \label{Introduction}
The large-scale structure (LSS) of the Universe consists of nodes, filaments, sheets, and voids, forming the cosmic web. These structures grew from small density fluctuations in the early Universe \citep{Zeld1970, Bond1996} and are reproduced in cosmological simulations such as the Millennium Simulation \citep{Springel2005} and IllustrisTNG \citep{Nelson2018}, which also provide a framework for connecting the cosmic web to the formation and evolution of galaxies \citep{Balogh2004, Kauffmann2004, Chartab2025}. Observable properties of galaxies, such as stellar mass and SFR, are known to be affected by their environment \citep{Peng2010, Scoville2013, Darvish2016, Chartab2020, Taamoli2024}. Although many features of the large-scale distribution of matter are successfully predicted by the standard $\Lambda$CDM model of cosmology, several observational tensions remain, including the Hubble tension (see, e.g., \citealt{DiValentino2021} for a review) and the $\sigma_{8}$ tension \citep{Hildebrandt2017, Joudaki2018}. $\Lambda$CDM also faces challenges on smaller, highly nonlinear scales, including discrepancies in halo structure, galaxy clustering, and sub-halo abundance \citep{Bullock2017,Perivolaropoulos2022,Sales2022,Khoshtinat2024}. These challenges highlight the importance of studying the cosmic web to understand galaxy evolution and constrain cosmological parameters.

Early spectroscopic surveys, such as Two-Degree Field Galaxy Redshift Survey (2dFGRS; \citealt{Colless2003}) and the Sloan Digital Sky Survey (SDSS; \citealt{York2000}), provided the first direct maps of this network on scales of hundreds of megaparsecs \citep{Doroshkevich2004, Tegmark2004}. Deep multi-wavelength programs, including COSMOS \citep{Scoville2007} and CANDELS \citep{Grogin2011, Koekemoer2011}, extended these studies to higher redshifts, enabling statistical measurements of how galaxy properties vary with environment across cosmic time. At low redshifts, dense regions host massive, quiescent galaxies, while less dense regions contain a higher fraction of star-forming systems \citep{Dressler1980, Balogh2004, Darvish2015}. With the advent of deep optical and near-infrared imaging surveys, it has become possible to extend environmental studies beyond the local Universe. However, probing higher redshifts ($z>3$) introduces new challenges, such as smaller sample sizes, larger photometric uncertainties, and incomplete coverage of low-mass systems. 

Stellar mass and SFR are the most fundamental quantities for understanding galaxy growth and quenching, as they directly trace both the buildup of stellar populations and ongoing star formation activity. Stellar mass records the cumulative history of star formation and mergers \citep{Gallazzi2005, Tremonti2004, Tacconi2020}, correlates with morphology and quiescence \citep{Kauffmann2003, Peng2010}, metallicity \citep{Tremonti2004}, and connects to structural parameters (such as size and mass-density) and halo properties that regulate star formation \citep{Shen2003, Dekel2006, Moster2010, Behroozi2013,vanDerWel2014, Gozaliasl2025, Yang2025}. Because stellar mass reflects the combined effects of internal physics and environmental processes, it is an essential parameter for tracing the imprint of these influences and for bridging galaxy-scale processes with the large-scale cosmic web. Moreover, SFR measures the current rate of stellar mass growth \citep{Kennicutt1998, Madau2014, Tacconi2020} and reflects feedback processes that regulate the gas reservoir and suppress subsequent star formation \citep{Hopkins2012, Naab2017, Maiolino2019}. Thus, investigating how SFR evolves across different environments helps clarify the role that external conditions play in galaxy evolution.

Part of the observed dependence of SFR on environment is driven by quenching, rather than changes in the SFR itself \citep{Taamoli2024,Chartab2020}. Both stellar mass and environment can control quenching \citep{Peng2010}. Internal mechanisms such as AGN feedback, morphological stabilization, and shock heating tend to suppress star formation in massive galaxies \citep{Dekel2006, Martig2009, Schawinski2014}. External mechanisms including ram-pressure stripping, strangulation, and galaxy interactions act in dense environments to remove or heat gas and halt star formation \citep{Gunn1972, Moore1999, Poggianti2017, Wetzel2013}. Recent studies show that mass and environmental quenching efficiencies both increase with stellar mass, with mass quenching dominating at the high-mass end and environmental effects becoming more important for low-mass systems, particularly at low redshift \citep{Chartab2020, Taamoli2025, Jin2024}. Therefore, stellar mass growth, ongoing star formation, and the quenching processes that shut down star formation are all connected to each other. Stellar mass traces the cumulative history of star formation, while the SFR reflects the current level of activity, and the specific star-formation rate ($\mathrm{sSFR} \equiv \mathrm{SFR}/M_\ast$) links the two by showing the relative pace of growth. Physical mechanisms behind quenching regulates how and when this growth is halted, and its strength depends on both internal factors related to mass and external influences from the environment. Understanding how these trends depend on environment is therefore a multidimensional problem, requiring all of these parameters to be considered simultaneously.

At high redshift, observational limitations led to mixed results in previous studies. Some studies find a weakening or flattening of SFR relation with density \citep{Scoville2013, Darvish2016}, others report no significant correlation \citep{Grutzbauch2011}, and some detect enhanced star formation in dense environments at $z \gtrsim 2$ \citep{Elbaz2007, Lemaux2022}. The release of COSMOS2020, with its deeper photometry and improved photometric redshifts, extended such measurements to $z \lesssim 4$ \citep{Weaver2022,Taamoli2024}. However, even with these advances, samples remain highly biased at the highest redshifts, particularly for low-mass galaxies, where photometric uncertainties and selection effects influence both the completeness and reliability of environmental trends \citep{Taamoli2024, Brinch2023}.

To overcome these depth and completeness limitations, COSMOS-Web was designed to provide the first wide-area, JWST-based view of the cosmic web, covering $0.54$ deg$^{2}$ of the COSMOS field with NIRCam imaging and complementary MIRI observations \citep{Casey2023}. Its area is sufficiently wide to capture the full diversity of the cosmic web, and its near-infrared depth (reaching $5\sigma$ depths of $27.2$–$28.1$ AB mag) pushes detections to $z \sim 10$ \citep{Casey2023, Casey2024, Franco2025, Shuntov2025}. COSMOS-Web provides a mass-selected sample, reaching 80\% completeness at $\sim10^{9}\,M_{\odot}$ at $z \sim 10$ \citep{Shuntov2025}. These advances improve the accuracy of large-scale structure reconstructions and extend environmental studies to higher redshifts than previously possible.

This paper focuses on the large-scale structures identified from the COSMOS-Web survey, the first wide-area, deep, near-infrared–selected survey designed to study the cosmic web and the impact of environment on galaxy properties. This paper is structured as follows. In Section \ref{sec:Data}, we describe our data set and sample selection. In Section \ref{sec:method}, we describe the method we used to measure density maps. In Section \ref{sec:results}, we present the density map results. In Section \ref{sec:Discussion}, we compare our results from COSMOS-Web with COSMOS2020, and then discuss our results on the evolution of galaxy properties with the cosmic web. Finally, we summarize our findings in Section \ref{sec:Summary}.

In this work, we adopt a flat $\Lambda$CDM cosmology with $H_{0} = 70 \,\mathrm{km\,s^{-1}\,Mpc^{-1}}$, $\Omega_{\mathrm{m},0} = 0.3$, and $\Omega_{\Lambda,0} = 0.7$. All magnitudes are expressed in the AB system \citep{Oke1974}, and a Chabrier initial mass function (IMF) is assumed \citep{Chabrier2003}.

\section{Data} \label{sec:Data}

\subsection{COSMOS-Web}
The COSMOS-Web survey \citep{Casey2023} provides deep JWST observations in the central $0.54\;\mathrm{deg}^2$ of COSMOS in four NIRCam bands (F115W, F150W, F277W, F444W), reaching a $5\sigma$ point-source depth of $27.5$–$28.2$ mag.  In addition, MIRI F770W covers $\sim0.20\;\mathrm{deg}^2$ with a $5\sigma$ depth of $25.3$–$26.0$ mag \citep{Wright2023}. Data reduction for NIRCam and MIRI are described in \citet{Franco2025} and \citet{Harish2025}, and the photometric catalog is presented in \citet{Shuntov2025}. The COSMOS-Web catalog contains about $780{,}000$ galaxies, with roughly $80\%$ completeness at $\log(M_{\star}/M_{\odot})\sim9$ by $z\sim10$ and $\log(M_{\star}/M_{\odot})\sim7$ by $z\sim0.2$, an improvement of about $1$ dex compared to COSMOS2020.  The combination of depth, area, and full multi-wavelength coverage makes COSMOS-Web the most powerful data set to study galaxy formation and evolution up to high redshifts ($z\sim7$).

\subsection{COSMOS2020}
To directly assess the improvements of COSMOS-Web in LSS studies relative to previous surveys, we use the COSMOS2020 catalog, which provides multi-wavelength photometry over the full $2\;\mathrm{deg}^2$ of the COSMOS field \citep{Weaver2022}. It combines imaging from over 30 bands spanning $0.3$–$8\;\mu$m, coming from a range of ground and space-based observatories. Optical imaging includes broad-band ($g, r, i, z, y$) data from Subaru/Hyper-SuprimeCam (HSC), 11 medium-band and 2 narrow-band filters from Subaru/Suprime-Cam that were originally obtained for COSMOS2015 and subsequently incorporated into COSMOS2020 \citep{Aihara2017, Taniguchi2007, Taniguchi2015, Laigle2016, Weaver2022}, and CFHT/MegaCam $u$-band imaging from the CLAUDS program \citep{Sawicki2019}. Near-infrared data are provided by UltraVISTA DR6 from VISTA/VIRCAM \citep{McCracken2012, Sutherland2015}. Space-based data include HST/ACS F814W \citep{Koekemoer2007, Koekemoer2011} and Spitzer/IRAC mid-infrared imaging \citep{Sanders2007,Ashby2018}.

The COSMOS2020 release provides two complementary catalogs, the ``CLASSIC'' catalog based on aperture photometry and the \texttt{Farmer} catalog produced with the \texttt{Farmer} profile-fitting tool \citep{Weaver2022}. The \texttt{Farmer} catalog measures total fluxes without aperture corrections and models blended sources more effectively, especially in low-resolution images such as IRAC. In contrast, \texttt{CLASSIC} aperture photometry tends to underestimate total flux and cannot model blended objects simultaneously. Although the photo-$z$ quality is similar between the two, \texttt{Farmer} performs better at fainter magnitudes ($i \ge 24$), which mainly correspond to high-redshift galaxies. For our density estimates, we use photometric redshifts and galaxy properties derived from the \texttt{Farmer} measurements using \texttt{LePhare} spectral energy distribution (SED) fitting \citep{Weaver2022}.

\subsection{Sample Selection}
\begin{figure}
    \centering
    \includegraphics[width=1\linewidth]{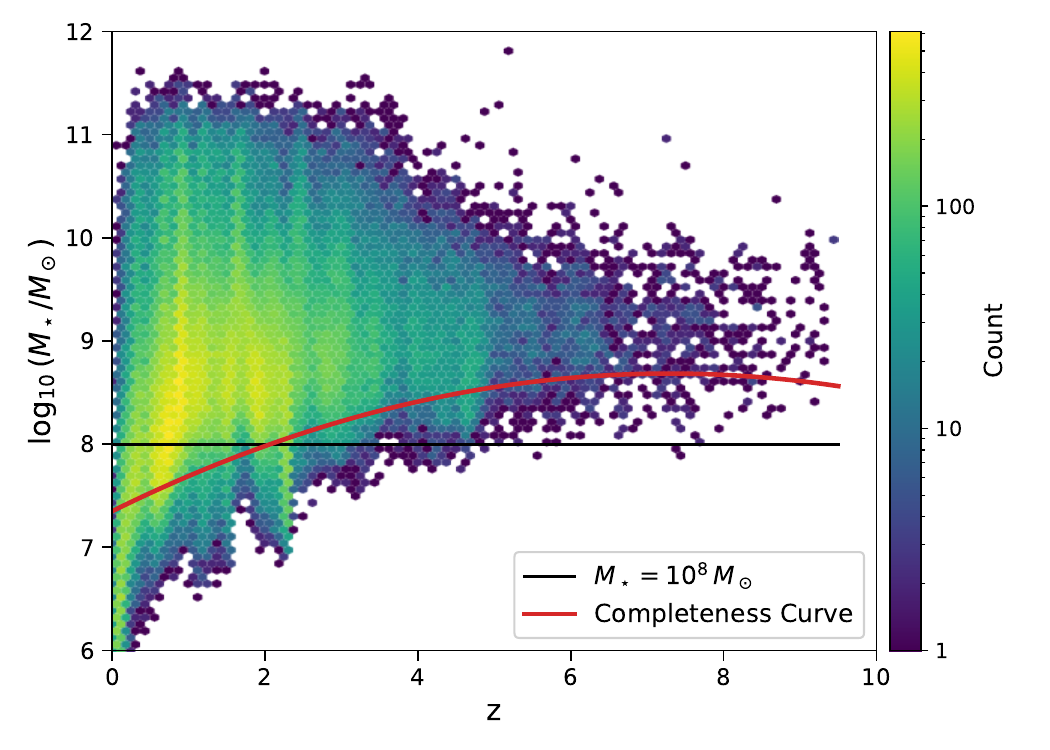}
    \caption{Galaxy stellar mass as a function of redshift. 
The red curve shows the stellar-mass completeness limit derived following the method of \citet{Pozzetti2010}, 
obtained by rescaling stellar masses to the total F444W magnitude limit of the survey ($27.5$) 
and taking the 90th percentile of the resulting distribution in each redshift bin. 
The solid red line represents the best-fit polynomial in $(1 + z)$. The black line marks the stellar mass threshold of $\log(M_\star/M_\odot) = 8$ used in our sample selection.}
    \label{fig:Mass_vs_zPDF}
\end{figure}

We define selection criteria for the COSMOS-Web and COSMOS2020 catalogs to construct density maps, ensuring the samples are as complete as possible and directly comparable for evaluating JWST’s improvements in large-scale structure mapping. At low redshift, our catalogs include many low-mass galaxies ($\log(M_\star/M_\odot) < 8$) that cannot be detected at higher redshift due to observational limits. Including these faint systems only at low $z$ would cause the density maps at different epochs to represent intrinsically different galaxy populations, introducing bias when tracing the evolution of the LSS and its connection to galaxy properties. Based on the COSMOS-Web catalog, we adopt a stellar mass threshold of $\log(M_\star/M_\odot) = 8$. This value provides a balance between completeness and uniformity at different redshifts. If we set the cut lower, the density maps at low $z$ would include numerous faint, low-mass galaxies that are undetectable at higher redshift, biasing the comparison of structures across cosmic time. Conversely, adopting a higher cut would remove a significant fraction of galaxies at low redshift, artificially diluting the density field and erasing some structures. We estimate the stellar-mass completeness limit for the COSMOS-Web catalog following the method of \citet{Pozzetti2010, Shuntov2025AA}, by rescaling stellar masses to the total F444W magnitude limit of the survey ($27.5$; \citealt{Shuntov2025}) and taking the 90th percentile of the resulting distribution in each redshift bin. The red curve in Figure~\ref{fig:Mass_vs_zPDF} shows this completeness function and the black line illustrates the mass cut. The sample remains complete down to $\log(M_\star/M_\odot)\sim8.5$ at $z\sim7$. A cut at $\log(M_\star/M_\odot) < 8$ therefore represents a reasonable compromise, as it minimizes redshift-dependent selection effects while still retaining enough galaxies to trace LSS consistently across all redshifts.

With a magnitude limit of $K_s=24.5$ (AB) on the COSMOS2020 catalog, the sample is complete to $i\approx24.9$ (AB), and $\sim32\%$ of  the sources lie in the range $24\le i\le24.9$ where \texttt{Farmer} performs better. The main limitation of \texttt{Farmer} is that it cannot model sources near bright objects, leading to the exclusion of $\sim16\%$ of sources in the star-masked areas. We discard all objects near stars, \texttt{STAR\_HSC} masked regions, to make sure the photometric redshifts are accurate and reliable \citep{Coupon2018}. We also exclude objects with bad spectral energy distribution (SED) fittings ($\chi^2/\text{Number of filters used}<3$) \citep{Casey2024}. We set a magnitude limit of $27.35$ for F150W for the COSMOS-Web catalog \citep{Toni2025GG}. In order to have reliable redshift estimates, we only use galaxies with $\sigma_z/(1+z)<0.1$ for both catalogs \citep{Taamoli2024}. After applying all these criteria, the COSMOS-Web remains with $\sim161,000$ galaxies, and the COSMOS2020 narrows down to $\sim79,000$ galaxies. The sample selection criteria are summarized in Table \ref{table:sample_selection}.

\begin{table*}[t!] % 't' for top placement; can also use 'h' or 'b'
\centering
\begin{tabular}{lcc}
\toprule
\textbf{Selection Criterion} & \textbf{COSMOS-Web} & \textbf{COSMOS2020} \\
Sky Area                    & COSMOS-Web $0.54$ deg$^2$ area & Limited to COSMOS-Web area (to avoid edge effects) \\
Magnitude Limit             & F150W $< 27.35$ & $K_s < 24.5$\\
Redshift Limit                   & $z<9.5$ & $z<6.0$ \\
Stellar Mass Cut $(M_\odot)$          & $10^8$ & $10^8$ \\
Photometric Redshift Quality & $\sigma_z/(1+z) < 0.1$ & $\sigma_z/(1+z) < 0.1$ \\
SED Fitting Quality         & $\chi^2/N_\mathrm{filters} < 3$ & $\chi^2/N_\mathrm{filters} < 3$ \\
Star Mask                   & \texttt{STAR\_HSC} excluded & \texttt{STAR\_HSC} excluded \\
\bottomrule
\end{tabular}
\caption{Selection thresholds and filtering criteria applied to COSMOS-Web and COSMOS2020 catalogs for density field construction.}
\label{table:sample_selection}
\end{table*}

\section{Density Estimation Method}
\label{sec:method}
The environmental density can be estimated using various methods, including weighted kernel density estimation (wKDE), weighted $k$-nearest neighbors, weighted Voronoi tessellation, and weighted Delaunay triangulation. Based on the comparison of these methods by \citet{Darvish2015}, wKDE and Voronoi tessellation best reproduce the density field in simulations, with wKDE showing higher stability against shot noise and sparse sampling. We reconstruct the environmental density field using the wKDE method developed by \citet{Darvish2015}, adapted and applied in recent studies \citep{Chartab2020, Taamoli2024, Brinch2023, Brinch2024}. This approach uses wKDE over photometric redshift slices, considering galaxy redshift uncertainties and correcting for survey geometry and masking. We use the same method for both catalogs and describe it below. 

\subsection{Redshift Slicing Based on Physical Scale}

We define redshift slices based on a fixed comoving thickness of $\Delta\chi = 35\,h^{-1}\,\mathrm{Mpc}$. This physical scale is chosen to be larger than the characteristic size of galaxy groups and clusters, and sufficiently wide to encompass redshift-space distortions (RSDs) and photometric redshift errors \citep{Muldrew2015}.

The redshift width $\Delta z$ for each slice is computed using:

\begin{equation}
\Delta z = \frac{H_0}{c} (1 + z) \sqrt{\Omega_{m,0} (1+z)^3 + \Omega_{\Lambda,0}} \times \Delta\chi.
\end{equation}

This results in 157 slices spanning $z = 0.4$ to $z = 9.5$, with $\Delta z$ varying from $\sim0.014$ at $z = 0.4$ to $\sim0.208$ at $z = 9.5$. However, density maps of COSMOS-Web can only be reliable up to $z\sim7$, which we will discuss in detail in Section \ref{sec:max_z}. The remaining slices are only considered as buffer for redshift probability distribution functions (PDFs), to have more reliable weights, and for more accurate density estimation at higher redshifts.

\subsection{Redshift Probability Weighting}

Each galaxy contributes to multiple redshift slices according to its photometric redshift probability distribution function (PDF), $P(z)$. The weight of a galaxy $g$ in slice $s$ is given by:

\begin{equation}
w^s_g = \int_{z_s^{\mathrm{min}}}^{z_s^{\mathrm{max}}} P_g(z)\, dz,
\end{equation}

where the integration is over the bounds of slice $s$. For galaxies with uni-modal PDFs, $P(z)$ is approximated as a Gaussian centered at the photo-$z$ with $\sigma_z$ equal to the 68\% confidence interval \citep{Laigle2016}. To reduce computation time, we apply a weight threshold $w_{\mathrm{th}} = 0.05$ recommended by \citet{Taamoli2024}, meaning galaxies contribute to a slice only if $w^s_g > 0.05$.

\subsection{Kernel Density Estimation with von Mises–Fisher Kernel}

We estimate the surface density field $\sigma^s(\mathbf{X})$ in each redshift slice $s$ using a weighted kernel density estimator:

\begin{equation}
\sigma^s(\mathbf{X}_i) = \sum_{g} w^s_g\, K(\mathbf{X}_i; \mathbf{X}_g, b_g),
\end{equation}

where $\mathbf{X}_i$ and $\mathbf{X}_g$ denote the angular coordinates of the target and contributing galaxies, $w^s_g$ is the redshift weight, and $b_g$ is the adaptive kernel bandwidth. The kernel function is chosen to be the von Mises–Fisher (vMF) distribution \citep{Chartab2020}, a spherical analog to the Gaussian:

\begin{equation}
K(\mathbf{X}_i; \mathbf{X}_g, b) = \frac{\exp(\cos{(\psi)}/b^2)}{4\pi b^2 \sinh(1/b^2)} ,
\end{equation}

where $\psi$ is the angular separation between $\mathbf{X}_i$, and $\mathbf{X}_g$. $b$ is the bandwidth of the vMF kernel, which shows how much a galaxy at $\mathbf{X}_g$ can affect the density at $\mathbf{X}_i$. This kernel naturally handles angular coordinates and becomes equivalent to a Gaussian for small $\psi$.

\subsection{Bandwidth Selection}

To balance smoothing and spatial resolution, we determine the optimal global bandwidth $b_s$ for each redshift slice using the Leave-One-Out Likelihood Cross-Validation (LCV) method \citep{Hall1982, Chartab2020}. This method tests different bandwidths and selects the one that best predicts the density at each galaxy's position when that galaxy is left out. LCV is given by:

\begin{equation}
\text{LCV}(b) = \sum_{k=1}^N \log \sigma_{-k}^s(\mathbf{X}_k),
\end{equation}

where $\sigma_{-k}^s$ excludes galaxy $k$ when computing the density at $\mathbf{X}_k$. The value of $b_s$ that maximizes this expression is chosen as the global bandwidth for that slice. 

After determining the bandwidth of each slice, $b_s$, we assign each galaxy its own adaptive bandwidth $b_g$ based on the local density. In dense regions, we reduce the bandwidth to capture finer structures; in sparse regions, we increase it to reduce noise.

\begin{equation}
b_g = b_s \left( \frac{\bar{\sigma}}{\sigma^s(\mathbf{X}_i)} \right)^\alpha,
\end{equation}

with $\bar{\sigma}$ the mean surface density in the slice and $\alpha = 0.5$ as recommended in \citet{Abramson1982}. This allows the smoothing scale to adjust based on the environment to improve the overall quality of the density field.

\subsection{Correction for Masked Regions}
In wide-field surveys such as COSMOS, bright stars and image artifacts contaminate large areas of the field, producing regions where photometry and SED fits are unreliable. These regions are masked and galaxies within them are excluded from the catalogs. If left uncorrected, such masks introduce systematic underdensities in the reconstructed density field, and large masks can mimic void-like structures that are not physically real. This effect becomes especially problematic when tracing large-scale structure, since density fluctuations on scales comparable to the mask size may be dominated by these observational artifacts rather than true cosmic variance.

\begin{figure}
    \centering
    \includegraphics[width=1\linewidth]{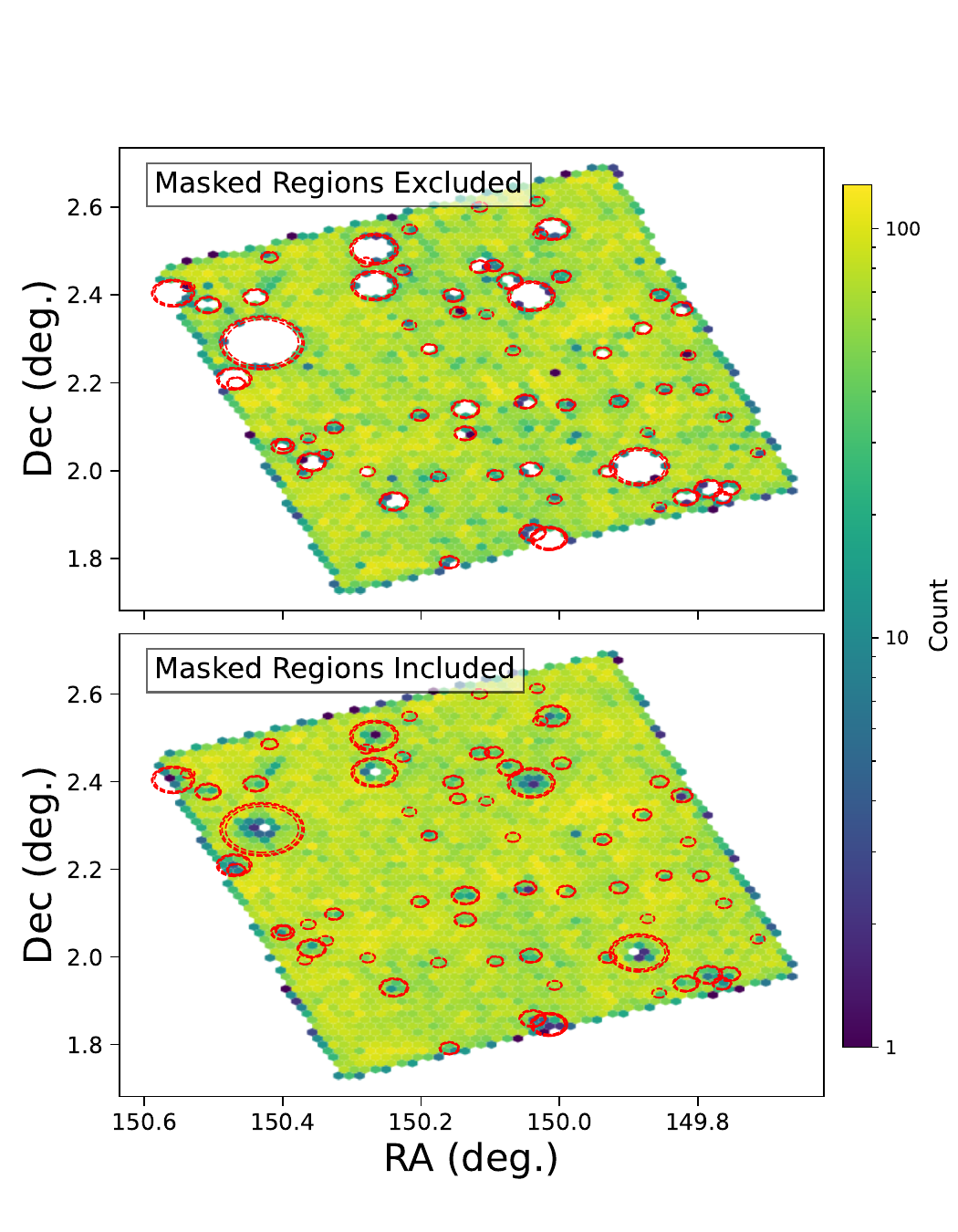}
\caption{Count of COSMOS-Web sources after all selection cuts. Red dashed circles show HSC-masked regions. Top: Masked regions excluded. Bottom: Masked regions included. Keeping detections inside the masks still leaves low number density inside the red circles, which would bias our LSS analysis.}
    \label{fig:masked_regions}
\end{figure}

Figure \ref{fig:masked_regions} shows the distribution of all sources in the COSMOS-Web field, one including all detected galaxies and one excluding those in the HSC mask. Even when galaxies within masked regions are included, their median detection rate is about $30\%$ lower, producing underdensities that resemble physical voids. This demonstrates the need to exclude sources in these regions and apply a correction to account for the masked regions.

To mitigate the effect of these regions, we follow the approach of \citet{Taamoli2024} and fill masked regions with a uniform distribution of artificial sources. For each redshift slice $s$, we assign a mean number density equal to the surface density of galaxies in the unmasked area of the field,  

\begin{equation}
\bar{n}_s = \frac{\sum_g w^s_g}{A_{\text{field}} - A_{\text{masked}}},
\end{equation}

where $A_{\text{field}}$ is the total field area and $A_{\text{masked}} = 0.127\,A_{\text{field}}$ is the masked area. This correction ensures that masked regions contribute neutrally to the density field, preventing spurious underdensities and yielding smoother and more physically consistent maps of LSS.

\subsection{Edge Correction}
Galaxy densities near the boundaries of the observed field are underestimated because part of the smoothing kernel extends outside the region covered by the catalog and therefore contributes no galaxies. To account for this, we calculate the effective fraction of the kernel that remains inside the observed field. For a galaxy at position $\mathbf{X}_i$, this fraction is  

\begin{equation}
\eta(\mathbf{X}_i) = \int_{\text{field}} K(\mathbf{X}_i; \mathbf{X})\, d\mathbf{X},
\end{equation}

where the integration is restricted to the observed field area. Physically, $\eta(\mathbf{X}_i)$ measures how much of the kernel centered at $\mathbf{X}_i$ lies within the data region. Well inside the field, the kernel is fully contained and $\eta(\mathbf{X}_i)\approx 1$. Near the edges, a portion of the kernel lies outside the field, giving $\eta(\mathbf{X}_i)<1$. Dividing the raw density by $\eta(\mathbf{X}_i)$ re-normalizes the estimate so that edge galaxies are treated consistently with those in the interior:  

\begin{equation}
\sigma^s_{\text{corrected}}(\mathbf{X}_i) = \frac{\sigma^s(\mathbf{X}_i)}{\eta(\mathbf{X}_i)}.
\end{equation}

This procedure corrects the systematic underestimation at field boundaries and yields reliable density estimates across the entire map \citep{Jones1993, Chartab2020}.

\subsection{Density Contrast Field}

We define the dimensionless density contrast as:

\begin{equation}
\delta = \frac{\sigma^s_{\text{corrected}}(\mathbf{X}_i) - \bar{\sigma}^s}{\bar{\sigma}^s},
\end{equation}

where $\bar{\sigma}^s$ is the mean surface density in the slice. The final result is a set of 157 angular density fields and corresponding density contrast maps spanning $z = 0.4$ to $9.5$.

\section{Results} \label{sec:results}
In this section, we present the results for each step of our method. We first show how galaxies are distributed across redshift slices defined by a fixed width of  $35\,h^{-1}\,\mathrm{Mpc}$. We then present the results for bandwidth selection, edge correction, and density fields in the COSMOS-Web field.  Lastly, we assess the highest redshift at which our density measurements remain reliable for the purpose of our work.

\subsection{Redshift Distribution}
\begin{figure*}
    \centering
    \includegraphics[width=1\linewidth]{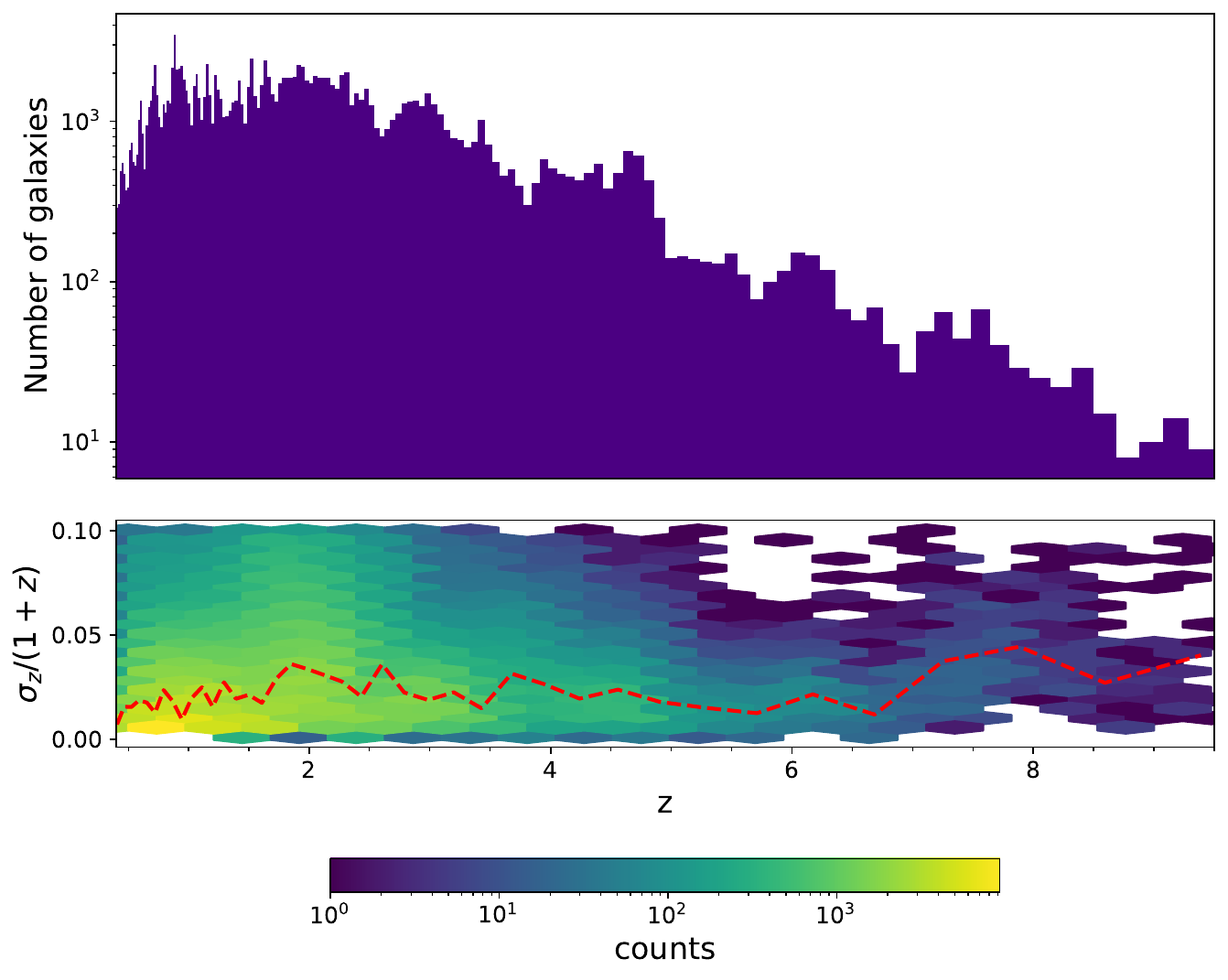}
    \caption{Top: Redshift distribution of galaxies in the COSMOS-Web sample. Each bin corresponds to a redshift slice used in the density field reconstruction.
Bottom: Normalized photometric redshift uncertainty, $\sigma_z / (1 + z)$, as a function of redshift. The color map shows the number of galaxies at each redshift–uncertainty bin on a logarithmic scale. The red dashed line shows the median uncertainty in each redshift slice.}
    \label{fig:redshift_distribution}
\end{figure*}

The distribution of galaxies in different redshift slices and their normalized photometric redshift uncertainties are shown in Fig. \ref{fig:redshift_distribution}. The top panel displays the histogram of galaxies as a function of redshift. Each redshift bin corresponds to a slice used for constructing the density maps Galaxies that contribute to each slice with a weight of $w^s_g > 0.05$ based on their redshift PDF are considered for building the density maps. At higher redshifts (e.g $z>6$), the number of galaxies is lower than later epochs due to observational limits.

Normalized redshift uncertainties remain relatively low at $z<6.5$ and gradually increase toward higher redshifts. While the median $\sigma_z / (1 + z)$ remains below typical quality thresholds across all redshifts, uncertainties grow at high redshift because of lower photometric signal-to-noise and the smaller number of filters in which galaxies are detected, which limits the constraints on their SEDs. This naturally impacts the precision of density estimates at those epochs.

Nevertheless, our approach incorporates the photometric redshift uncertainty for each galaxy, ensuring that the density field reconstruction reflects the underlying redshift uncertainties. This enables the detection of LSS even at high redshift, though with decreasing spatial resolution and greater uncertainty.

\subsection{Density Field Maps}
\begin{figure*}
    \centering
    \includegraphics[width=1\linewidth]{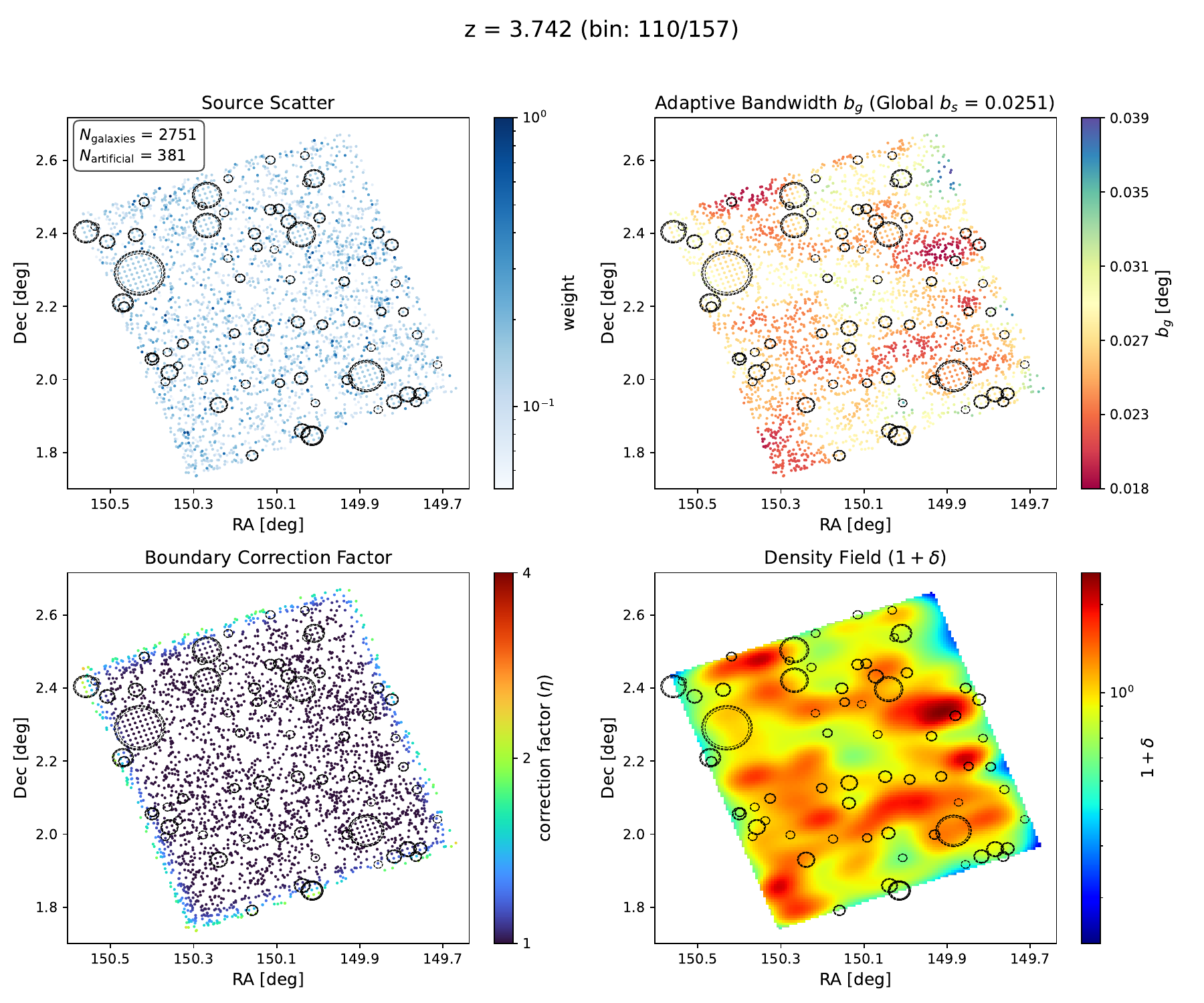}
    \caption{Steps in the construction of the density field for a redshift slice at $z \sim 3.7$. 
    Top left: distribution of galaxies and artificial sources colored by their weight. 
    Top right: adaptive bandwidth assigned to each galaxy. 
    Bottom left: boundary correction factor $\eta$ applied near survey edges. 
    Bottom right: final density field $1+\delta$. Circles show the HSC masked regions.}
    \label{fig:density_map_summary}
\end{figure*}

We visualize the step-by-step construction of the density field in Figure \ref{fig:density_map_summary} for one representative redshift slice at $z \sim 3.7$. Each panel demonstrates a component of the wKDE method explained in Section \ref{sec:method}. These intermediate maps allow us to evaluate the quality of the inputs and better understand the reliability of the final density construction.

In the top-left panel, we show the spatial distribution of galaxies within the slice, color-coded by their redshift weight contributions $w_g^s$. Masked regions (\texttt{STAR\_HSC}) have been filled by the average density of the field to avoid introducing artificial underdensities near bright stars. These added objects have the mean weight value of all the galaxies in the slice.

The top-right panel shows the variation in the adaptive kernel bandwidth $b_i$. Smaller bandwidths in dense regions allow for finer resolution, and larger bandwidths in underdense areas give smoother estimates to reduce noise. This adaptivity is important to resolve both compact overdensities and extended filamentary features. 

The bottom-left panel displays the boundary correction factor $\eta$, which accounts for edge effects and ensures that density values near the boundary regions are not underestimated due to truncated kernels. In the center of the field, where the full kernel fits inside the survey area, $\eta$ is close to $1$. Near the edges, where only part of the kernel overlaps with the data, $\eta$ increases to around $2$. In the corners, where most of the kernel is outside the field, $\eta$ can approach $4$.

\begin{figure*}
    \centering
    \includegraphics[width=\textwidth,height=\textheight,keepaspectratio]{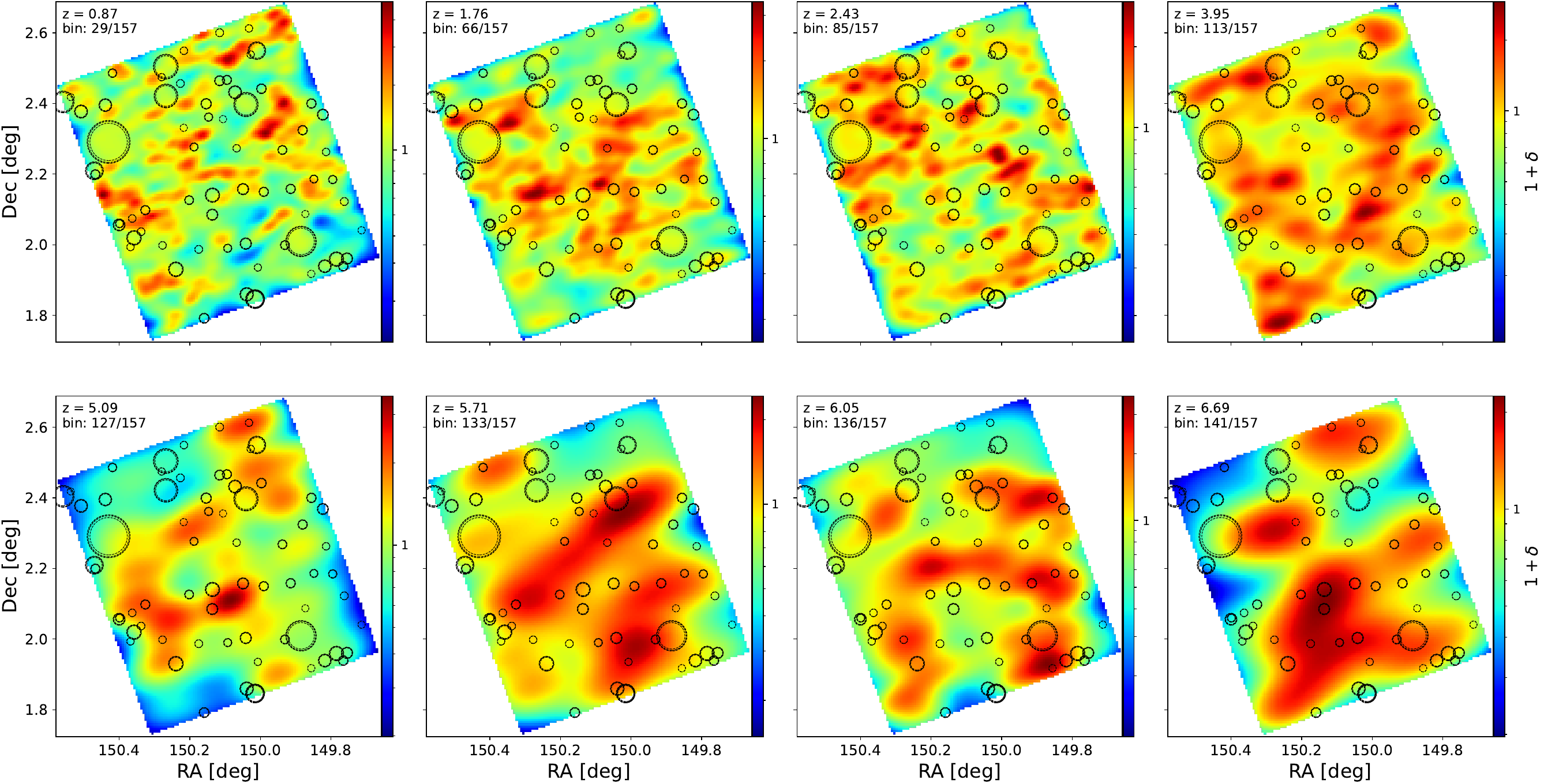}
    \caption{Evolution of overdensity maps in the COSMOS-Web field across cosmic time. Each panel shows the overdensity field for a fixed comoving slice; color bars are scaled independently in each redshift bin. While absolute overdensity amplitudes cannot be directly compared across panels, the maps illustrate the changing topology and distribution of LSS with increasing redshift. Circles mark the HSC star-masked regions.}
    \label{fig:density_grid}
\end{figure*}

Finally, the bottom-right presents the output density field. It is visualized with the dimensionless overdensity $1+\delta$. These maps are generated for all redshift slices and we use them to assign a density contrast value to each galaxy in the field. Figure \ref{fig:density_grid} shows the evolution of these 2D density maps with redshift in the COSMOS-Web field. Together, these figures show that the wKDE method recovers coherent structure even at early times, which allow us to connect galaxy properties to their environments over most of cosmic history.

\subsection{Density Contrast of Galaxies}

\begin{figure}
    \centering
    \includegraphics[width=1\linewidth]{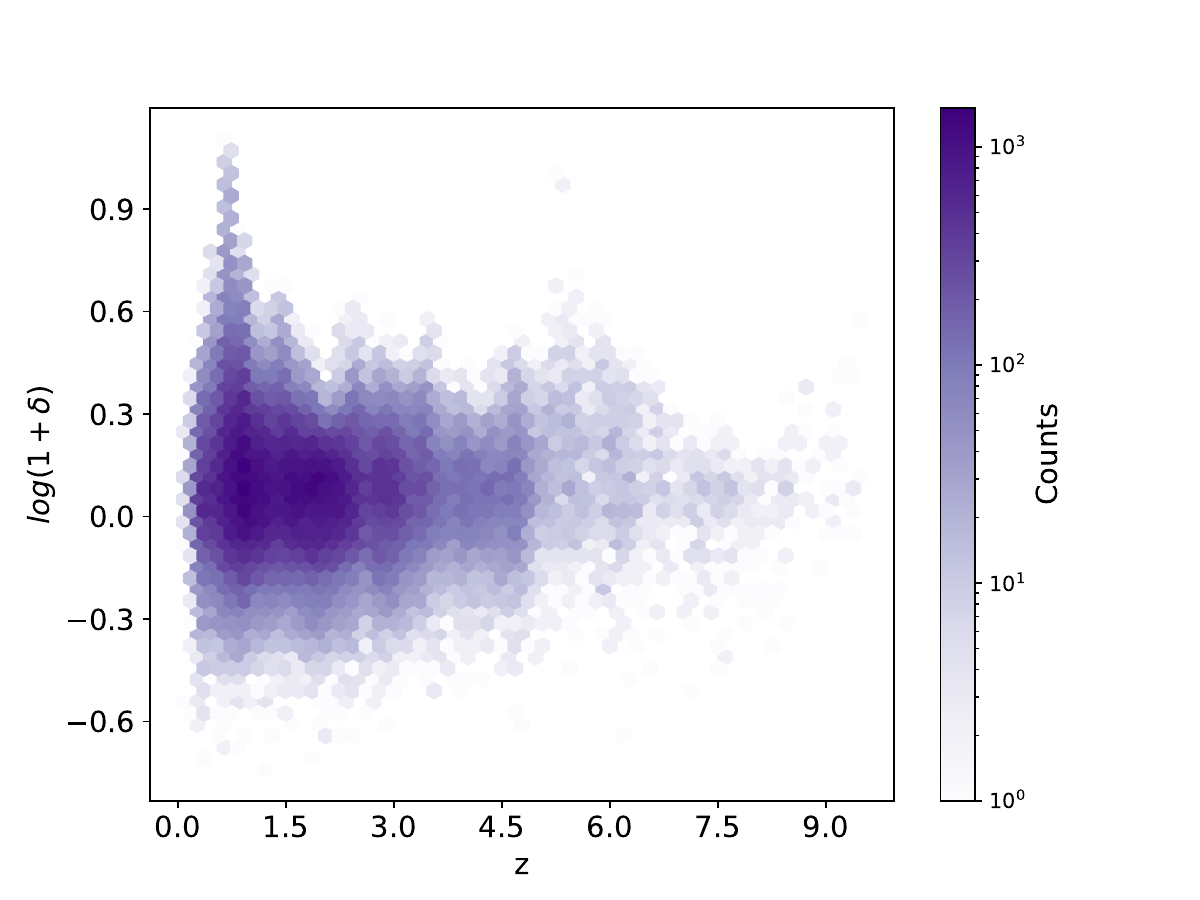}
    \caption{Distribution of galaxy overdensity, $\log(1+\delta)$, as a function of redshift in COSMOS-Web. Colors show the number of sources in each bin. The scatter of $\log(1+\delta)$ is broad at $z\lesssim2$, spanning low underdensities to high overdensities. The distribution narrows as redshift increases. Near $z\sim5$, the low-density tail compresses and galaxies occur more often in overdense regions, and by $z\sim7$ the high-density end is suppressed.}
    \label{fig:dist_overdensity-z}
\end{figure}

After generating overdensity maps for each slice in the COSMOS-Web, we find the overdensity associated with each galaxy in the field. Figure \ref{fig:dist_overdensity-z} shows the distribution of $\log(1+\delta)$ as a function of redshift. At lower redshifts ($z \lesssim 2$), galaxies span a wider range of overdensities. They vary from very low underdensities to strong overdensities. This is expected, as large-scale structure is more evolved at later times, with well-formed clusters and filaments becoming more prominent.

As redshift increases, the number of galaxies decreases, and the overall spread in $\log(1 + \delta)$ narrows. This is consistent with both the lower number density of galaxies in the early universe and the increased difficulty of resolving the structures. However, a significant range in density contrast remains detectable even at higher redshifts, with galaxies present in both relatively overdense and underdense regions. The decrease in counts at high redshifts is due to intrinsic rarity of detectable galaxies at early epochs. However, our selection criteria ensure reliable redshifts and SED fits in this range.

At lower redshifts, overdense regions typically correspond to matured environments such as clusters and filaments. These are epochs during which galaxies experience strong environmental effects. In contrast, at higher redshifts, the same values of $\log(1 + \delta)$ do not imply the same level of structural maturity. Overdense regions at early times are more likely to be associated with forming structures such as proto-clusters or early filaments that have not fully collapsed. This context is important for interpreting how galaxy properties relate to environment across redshift. In our two dimensional analysis of the cosmic web, filaments that are elongated in the line of sight can appear like clusters, and edge-on cosmic walls may appear as filaments due to projection. Additionally, photometric redshift estimates can introduce artificial accumulations at certain redshifts, leading to spurious enhancements or depletion in the apparent galaxy number density. Such effects are modest, but should be kept in mind when interpreting the redshift–overdensity distribution.

\begin{figure*}
    \centering    \includegraphics[width=\textwidth,height=\textheight,keepaspectratio]{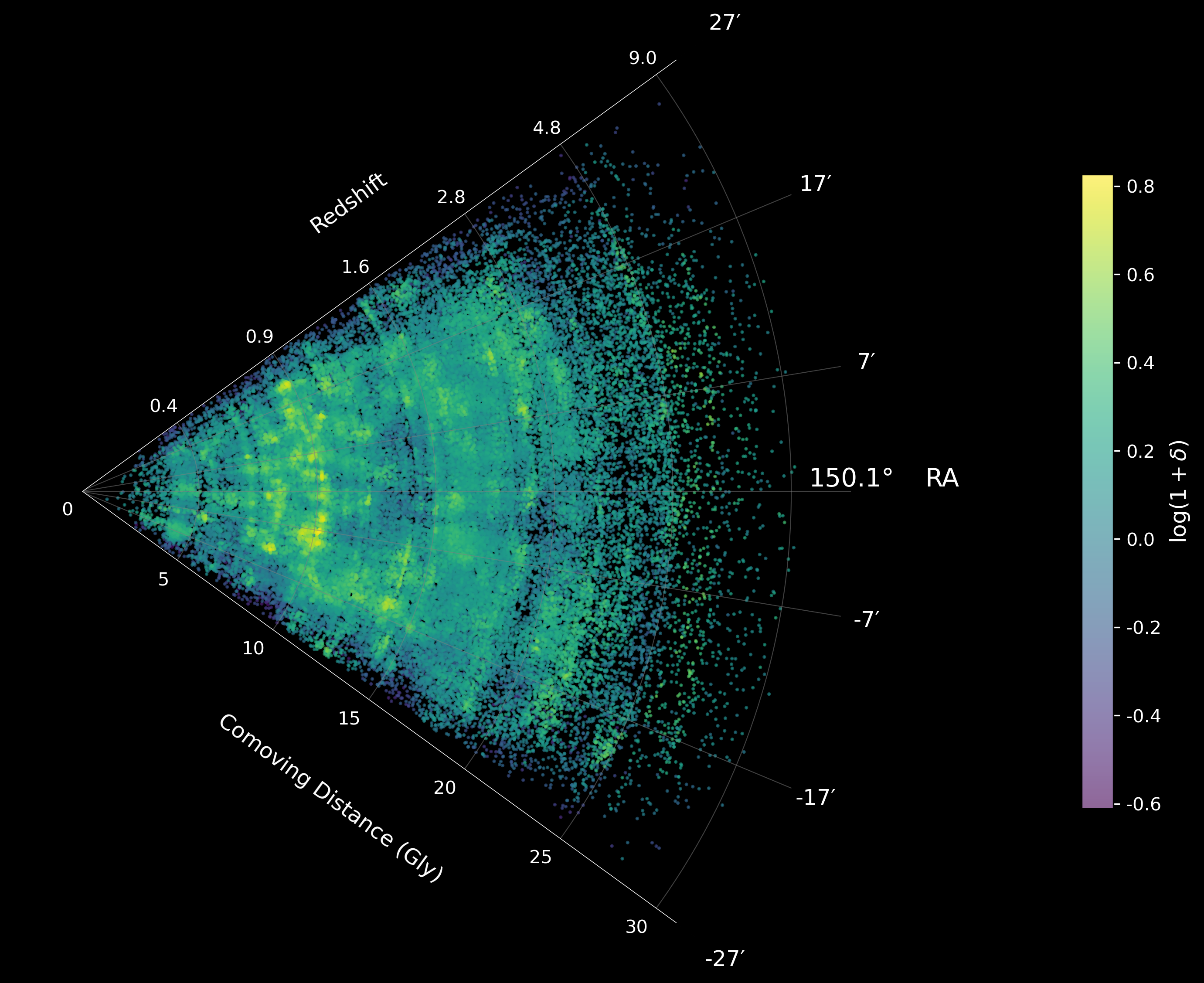}
    \caption{A slice of COSMOS-Web showing the large-scale structure of galaxies. 
    The plot selects galaxies within a $\Delta\mathrm{Dec} \sim 0.5^{\circ}$ band centered on the field median.
    The lower and upper radial coordinate gives the comoving distance in Gly and redshift, and the angular extent corresponds to RA offsets of $\pm 27'$ from the field center at $\mathrm{RA}=150.1^{\circ}$. 
    Each point is color-coded by local overdensity, expressed as $\log(1+\delta)$.}
    \label{fig:pizza}
\end{figure*}

To provide a complementary view, Figure \ref{fig:pizza} shows a thin slice of the COSMOS-Web in comoving coordinates. This diagram shows how galaxies trace large-scale structures over cosmic time, from the nearby Universe out to the highest redshifts probed. Galaxies in overdense regions are in filaments, groups, clusters, or proto-clusters, while those in underdense regions populate voids, providing the framework in which their physical properties (such as stellar mass, SFR, or sSFR) can be studied as a function of environment. These overdensity maps in the COSMOS-Web field thus serve as a powerful tool to investigate how galaxies evolve across environments, and how early large-scale structure may influence stellar mass growth and star formation activity, which will be discussed in detail in Section \ref{sec:Discussion}

\subsection{Maximum Reliable Redshift} \label{sec:max_z}
The maximum reliable redshift depends on the scientific context. Whether the density fields are sufficient for studying correlations between galaxy properties and large-scale environments depends on the intended study. This is inherently a more qualitative threshold, but coherence of density maps with adaptive bandwidth and statistical trends in overdensity values are important. Also, sample selection can affect the density fields specially at higher redshifts. Due to our selection criteria, $\Delta z / (1 + z) < 0.1$, and high-quality SED fits, the redshift probability distributions remain well constrained even at high redshifts. However, the decline in galaxy count can make density maps  inaccurate.

In Fig. \ref{fig:dist_overdensity-z}, as redshift increases, the distribution of $\log(1+\delta)$ gradually narrows. Around $z\sim5$, the low-density tail compresses, and galaxies are found more frequently in overdense regions. Then, by $z\sim7$, we observe the high-density end becomes suppressed, and detected galaxies in our sample mostly occupy intermediate-density regions. This shift may be partially affected by a bias in the sample at higher redshifts due to reionization, but likely it may reflect one of the early growth stage of large-scale structure. These trends show that while environmental structure exists at high redshift, the contrast between overdense and underdense regions diminishes, which limits how precisely we can define a galaxy’s environment.

We also visually inspect the density fields at high redshifts to determine up to what redshift they remain useful for statistical studies. Even though the number of galaxies decreases, the maps still display some structures and smooth variations in density. While the contrast between dense and sparse regions is lower than at lower redshifts, the maps remain usable for statistical studies of how galaxy properties relate to environment. Based on this, we adopt $z \sim 7$ as the maximum redshift for reliable density measurements for our purpose of study, which focuses on the connection between environment and galaxy properties such as stellar mass and star formation rate. At $z>7$, the density maps either become flat or exhibit sharp, non-smooth variations.

It is important to note that the interpretation of overdensities and underdensities changes with cosmic time. At lower redshifts, large-scale structure overdensities typically correspond to well-defined, gravitationally bound structures such as clusters, filaments, and walls. At higher redshifts ($z > 1.5$), overdense regions are usually associated with structures that are still in the process of formation, such as proto-clusters or early filaments and walls that have not yet reached their full structural development \citep{Muldrew2015, Chiang2013}.

\section{Discussion} \label{sec:Discussion}
Our density construction based on COSMOS-Web provides a new perspective on how galaxy environments evolve across cosmic time. In this section, we first discuss how COSMOS-Web improves measurements in the cosmic web compared to COSMOS2020. We then investigate how density contrast correlates with galaxy properties.

\subsection{Large-Scale Structure with COSMOS-Web Compared to COSMOS2020}

Accurately measuring the cosmic environment of galaxies relies heavily on the quality of photometric redshift estimates. Since our density fields are constructed using the redshift probability distributions of galaxies, more precise redshifts lead to more localized and reliable weight assignments within each slice. With COSMOS-Web, the precision of photometric redshifts has improved compared to COSMOS2020, with COSMOS-Web galaxies having a lower $\sigma_{\text{NMAD}}$ and reduced outlier fractions compared to COSMOS2020. These improvements are more significant for fainter galaxies \citep{Shuntov2025}. In COSMOS2020, broader and less accurate redshift PDFs cause galaxies' weight to spread to more slices, diluting any potential structure. In COSMOS-Web, narrower PDFs allow galaxies to contribute more accurately to the correct slice, resulting in cleaner and better resolved density maps.

\begin{figure*}
    \centering
    \includegraphics[width=1\linewidth]{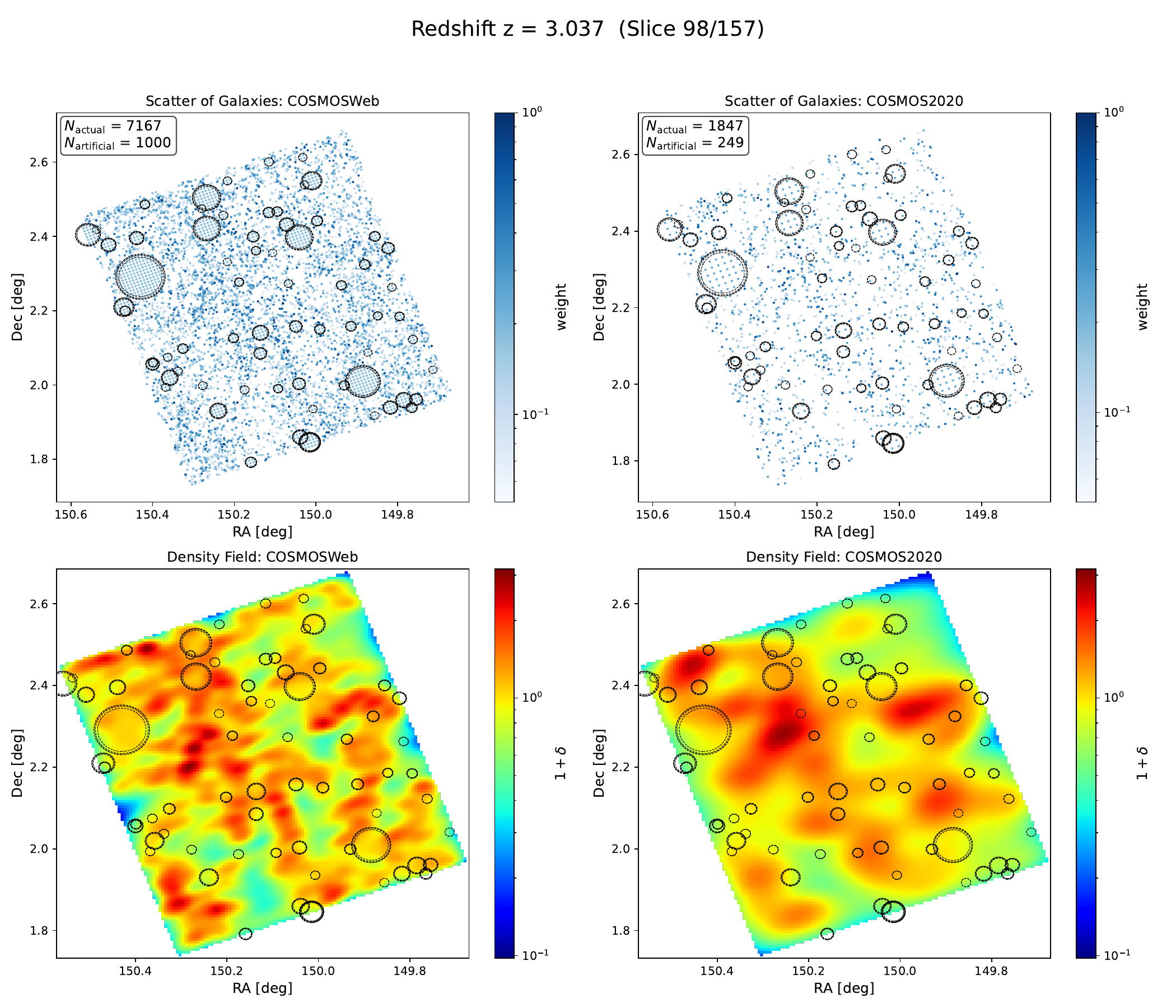}
    \caption{Comparison of density field reconstructions in a slice at $z \sim 3.0$ using COSMOS-Web (left) and COSMOS2020 (right). \textit{Top panels:} galaxy distributions in each slice, with points colored by their weights. \textit{Bottom panels:} corresponding overdensity maps, colored by $\log(1+\delta)$. The COSMOS-Web reconstruction recovers smoother and more coherent large-scale structures due to its deeper completeness, higher source density, and improved photometric redshift precision, whereas the COSMOS2020 map appears sparser and more diffuse, and missing some details in the structures. Circles indicate the HSC masked regions.}
    \label{fig:COSMOSWebvsCOSMOS2020}
\end{figure*}

In addition to better redshift precision, COSMOS-Web includes more galaxies than COSMOS2020 within the same area, especially at fainter magnitudes. Given the near-infrared selection criterion and deeper photometric data of COSMOS-Web, it contains more galaxies, especially at fainter limits, and therefore provides a more complete sample at all redshifts. The lower number of galaxies in COSMOS2020 can lead to gaps in coverage and more variation in the density maps, especially at higher redshifts. We limited the COSMOS2020 sample to the COSMOS-Web area, therefore maps are directly comparable.  These differences are visible in Figure~\ref{fig:COSMOSWebvsCOSMOS2020}, which compares one redshift slice ($z \sim 3$) between the two catalogs. In the top panels, the distribution of galaxies shows that COSMOS-Web includes far more galaxies in the slice. The bottom panel shows the density maps for both catalogs. The COSMOS2020 map appears smoother and more diffuse, with some structures entirely missed. This comparison illustrates how increased galaxy counts and improved redshifts together allow the detection of large-scale structure features that were not revealed before.

\begin{figure*}
    \centering
    \includegraphics[width=1\linewidth]{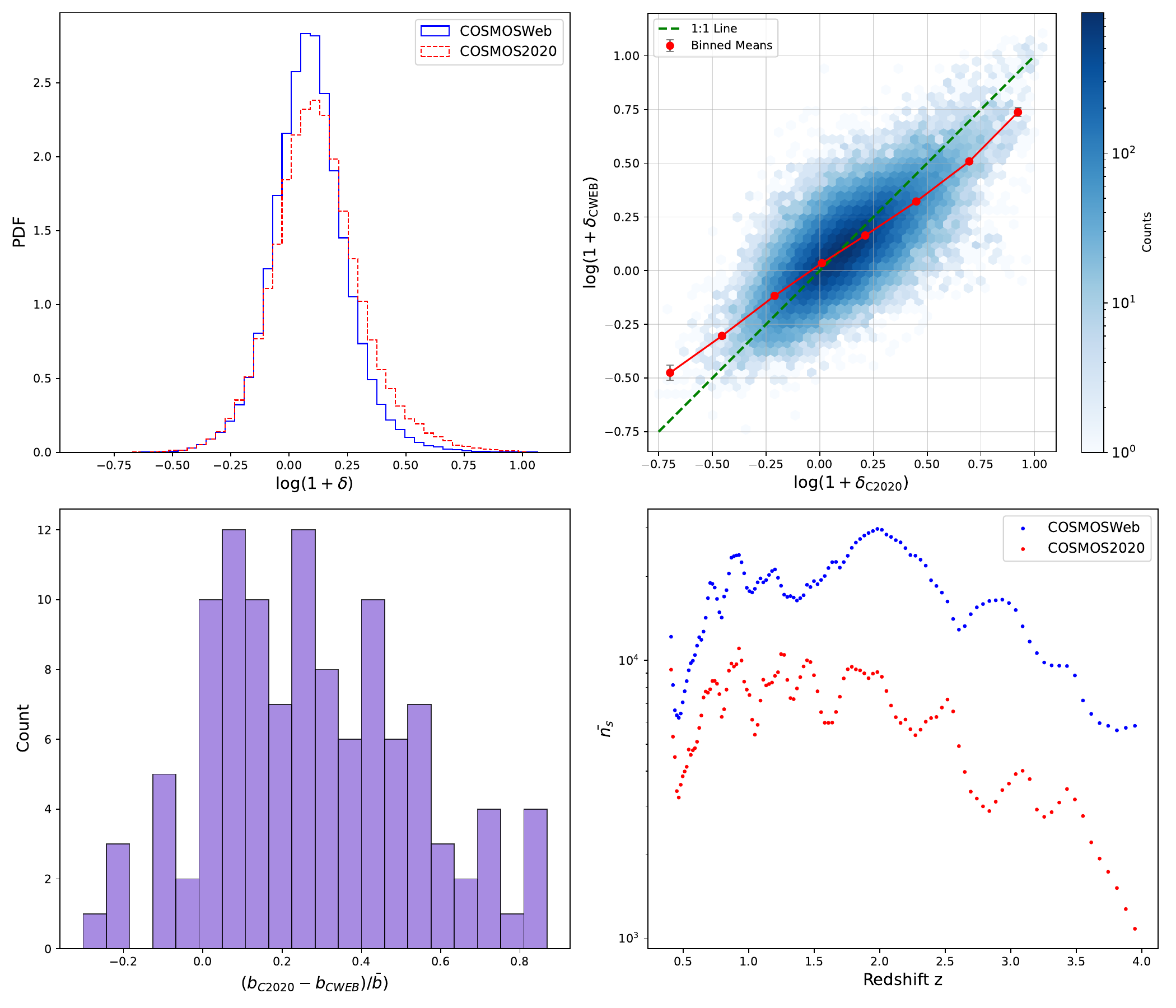}
    \caption{Comparison of density field diagnostics between COSMOS-Web and COSMOS2020 up to a redshift of $4$. 
    \textit{Top-left:} PDF of $\log(1+\delta)$ for both surveys. 
    \textit{Top-right:} galaxy-by-galaxy comparison of $\log(1+\delta_{\mathrm{CWEB}})$ and $\log(1+\delta_{\mathrm{C2020}})$ with binned means and 1:1 line (where the two quantities are equal). 
    \textit{Bottom-left:} relative difference in the global bandwidth $b_s$ for each redshift slice, normalized by their mean. 
    \textit{Bottom-right:} mean surface number density $n_s$ per slice for COSMOS-Web and COSMOS2020.}
    \label{fig:Cweb_C2020_comparison}
\end{figure*}

The comparison between COSMOS-Web and COSMOS2020 density contrasts up to $z\sim4$ reveals that the two catalogs differ systematically in how they represent both overdense and underdense regions. Figure~\ref{fig:Cweb_C2020_comparison} summarizes the key diagnostics:

\begin{itemize}

    \item The top-left panel shows the probability distribution functions of $\log(1+\delta)$ for all galaxies that are cross-matched between the two surveys. COSMOS2020 exhibits a broader high-density tail, indicating more regions assigned very large overdensities. COSMOS-Web, while recovering the same large-scale structures, produces a more concentrated distribution with fewer extreme peaks. This suggests that COSMOS2020 is prone to inflating the contrast in high-density regions.

    \item The top-right panel compares $\log(1+\delta_{\mathrm{CWEB}})$ and $\log(1+\delta_{\mathrm{C2020}})$ for individual galaxies cross matched between catalogs, with binned means overplotted. The relation follows the 1:1 line at intermediate densities but diverges at both ends. In high-density regions, density contrast in COSMOS2020 systematically exceeds COSMOS-Web, while in low-density regions, it produces less negative contrasts, making voids appear shallower compared to COSMOS-Web. The difference is caused by the differences in sampling, KDE smoothing ($b_s$), and density normalization ($\bar n_s$) between the catalogs.

   \item The bottom-left panel shows the relative difference in the optimal global bandwidth $b$ between COSMOS2020 and COSMOS-Web for each slice, normalized by the slice mean $\bar{b}$. In almost all slices, $b_{\mathrm{C2020}}$ is larger, reflecting the sparser sampling in COSMOS2020. With fewer tracers, the LCV algorithm selects a broader smoothing scale, which spreads galaxy signal into nearby voids and makes the density contrast less negative in those regions. COSMOS-Web, with denser sampling, adopts smaller $b$ values that preserve the amplitude of underdensities and finer features in structures. In overdense regions, the larger $b$ in COSMOS2020 also spreads clustered galaxy signal into surrounding cells, inflating peak contrasts.

    \item Finally, the bottom-right panel shows the mean number density $\bar{n}_s$ of galaxies per slice for the two surveys. COSMOS-Web has a consistently higher $\bar{n}_s$ across all redshifts, owing to its deeper flux limit and inclusion of faint galaxies. Since $\delta$ is defined relative to $\bar{\sigma}$, a higher mean surface density reduces the measured $(1+\delta)$ for any fixed absolute density $\sigma$. This normalization effect compresses the dynamic range of $\delta_{\mathrm{CWEB}}$ compared to $\delta_{\mathrm{C2020}}$, even when both are sampling the same underlying matter peaks. Also, the mean number density drop down significantly at higher redshifts in the COSMOS2020 sample ($z>3.5$).

\end{itemize}

All these diagnostics show that COSMOS2020 tends to overestimate densities in the most overdense environments and to underestimate the depth of underdense regions. COSMOS-Web, with its higher completeness, smaller smoothing bandwidths, and lower tracer bias, produces a more accurate mapping of the large-scale structure, even if the resulting $\delta$ field has a smaller dynamic range. This shift has direct implications for environmental studies. Trends measured with $\delta_{\mathrm{C2020}}$ may appear artificially steep in the highest-density bins and artificially shallow in the lowest-density bins, whereas $\delta_{\mathrm{CWEB}}$ better preserves the true relative contrast across environments.

\subsection{Evolution of Mass With Large-Scale Environment Density} \label{sec:mass_density}
\begin{figure*}
    \centering
    \includegraphics[width=1\linewidth]{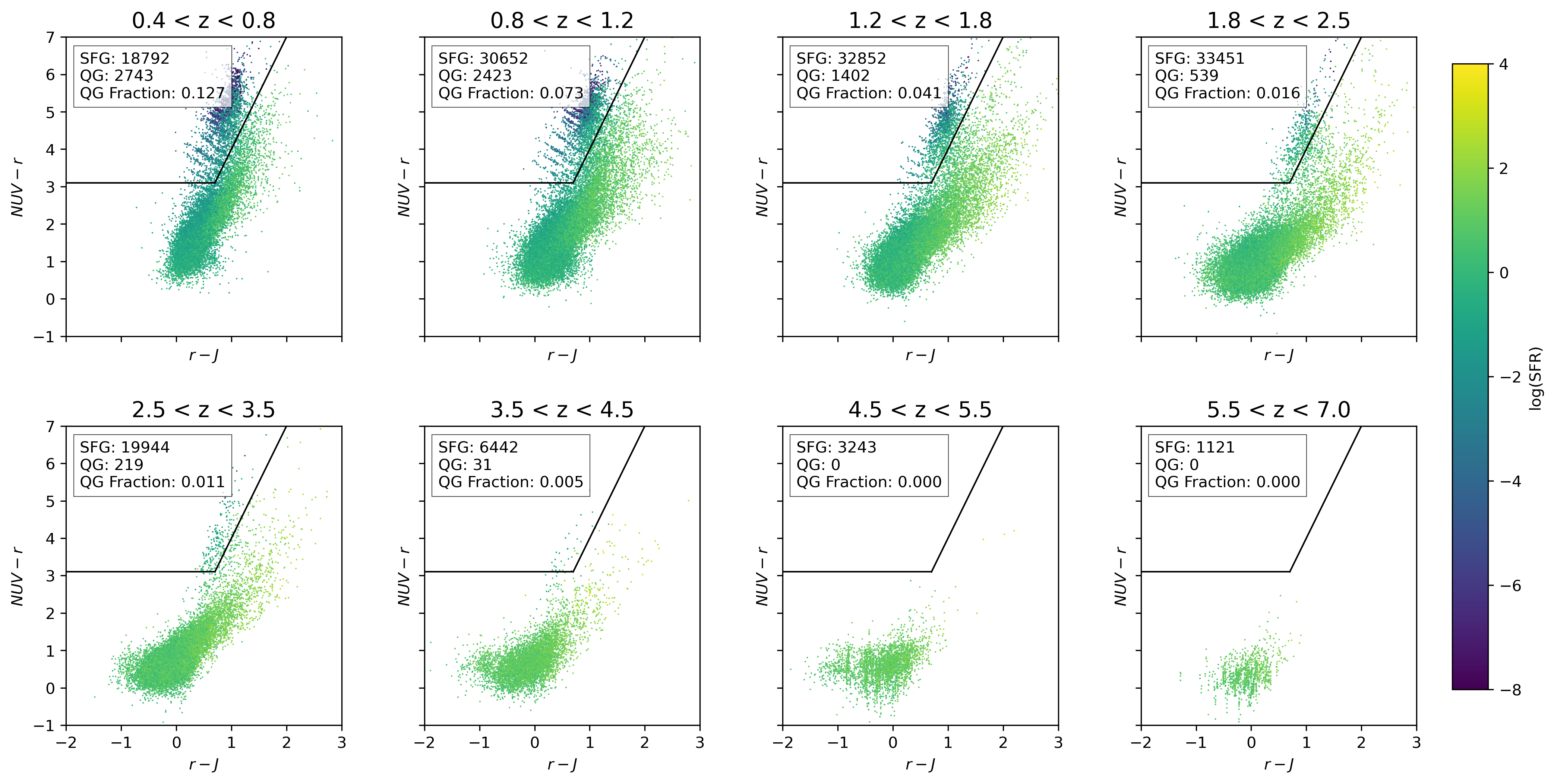}
    \caption{Rest-frame color-color diagrams ($NUV-r$ vs. $r-J$) in different redshift bins, used to separate SFGs from QGs. The color scale shows $\log(\mathrm{SFR})$. Each panel shows one redshift interval, with the black lines indicating the quiescent and star-forming selection criteria. The number of SFGs, QGs, and the quiescent fraction are reported in each bin.}
    \label{fig:qgs-sfgs-classification}
\end{figure*}

Mass is one of the most fundamental properties that reveals the evolution of galaxies. Understanding the environmental dependence of stellar mass as a function of cosmic epoch would give us insight in the role of LSS in driving the evolution of galaxies. Many previous studies have shown that, at low redshift, more massive galaxies reside in denser regions,  though its strength and statistical significance tend to decline with increasing redshift \citep{Darvish2015, Chartab2020, Lemaux2022, Taamoli2025}.  To better understand environmental effects on galaxy evolution, we analyze three samples, one with all galaxies (star-forming and quiescent), one with only star-forming galaxies (SFGs), and one with only quiescent galaxies (QGs). We identify QGs in each bin using the rest-frame color-color diagram (\textit{NUV}-$r$ vs.\ $r$-$J$) and the classification criteria from \citet{Ilber2013}. Galaxies with $\textit{NUV}-r > 3.1$ and $\textit{NUV}-r > 3(r - J) + 1$ are classified as quiescent. Figure \ref{fig:qgs-sfgs-classification} shows the distribution of galaxies in the color–color plane, showing QG and SFG populations, their SFR, and the corresponding quiescent fraction. We do not impose a threshold for SFR or sSFR when selecting our QG sample, as doing so could introduce biases in the subsequent analysis of SFR and sSFR versus density. Quiescent galaxy fraction decreases as we look at higher redshift, which is expected from the growth of quiescent population over cosmic time \citep{Ilber2013, Muzzin2013}. Also at redshifts $z > 2.5$, the QG sample is too limited for a robust physical interpretation.

\begin{figure*}
    \centering
    \includegraphics[width=1\linewidth]{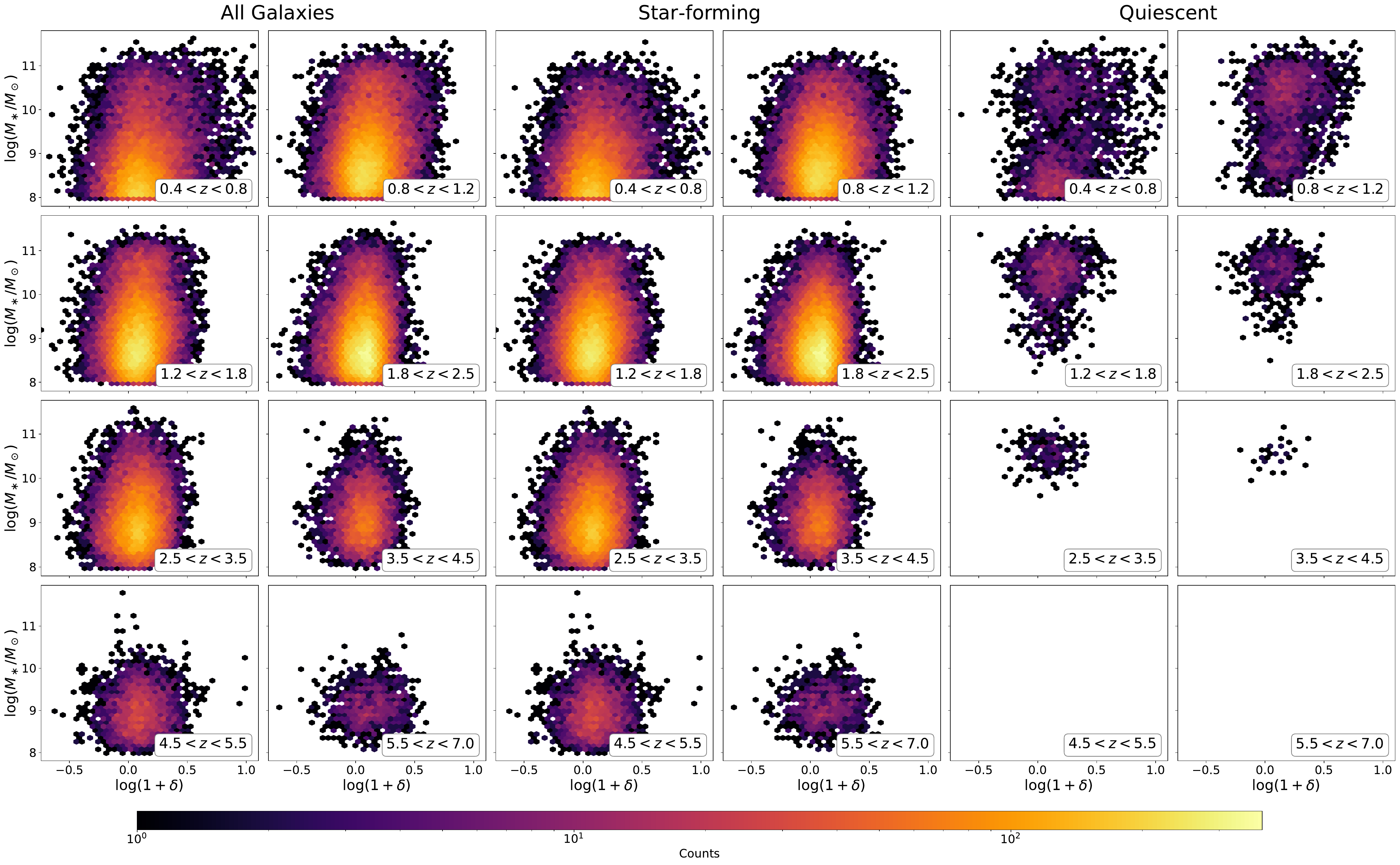}
    \caption{Stellar mass as a function of environmental overdensity in COSMOS-Web. 
    Each row shows a different redshift bin, increasing from top to bottom, and each column corresponds to a galaxy population: all galaxies (left), star-forming galaxies (middle), and quiescent galaxies (right). 
    Points are colored by number counts. 
    The distribution narrows with increasing redshift.}
    \label{fig:scatter_mass_vs_density}
\end{figure*}

In Figure~\ref{fig:scatter_mass_vs_density}, we show the distribution of stellar mass as a function of environmental overdensity in eight redshift bins from $z = 0.4$ to $z = 7$. At lower redshifts, a large number of galaxies are present across a wide range of environments. While most galaxies lie in the low-mass regime ($\log(M_*/M_\odot) \sim 8.5$–$10$), there is a population of higher-mass galaxies extending toward higher overdensities. As redshift increases, the distribution narrows both in stellar mass and in environmental density.

\begin{figure*}
    \centering
    \includegraphics[width=1\linewidth]{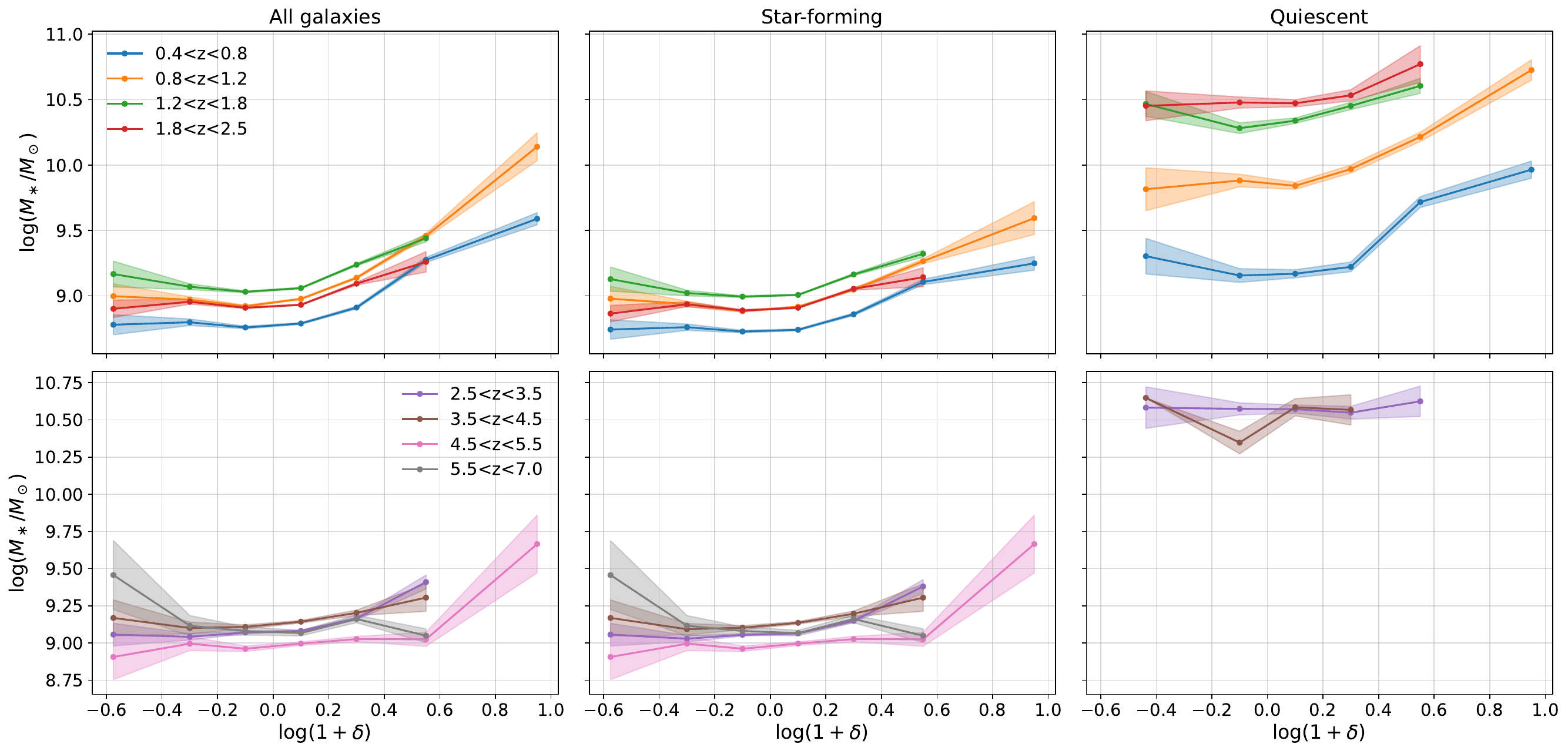}
    \caption{Average stellar mass as a function of environmental overdensity in COSMOS-Web. 
    Columns show all galaxies (left), star-forming galaxies (middle), and quiescent galaxies (right). 
    The top row corresponds to redshift bins in the range $0.4 < z < 2.5$, and the bottom row to $2.5 < z < 7.0$. 
    At $z \lesssim 2.5$, stellar mass increases with overdensity for all galaxies, most strongly for quiescent systems, while the trend for star-forming galaxies is weaker. 
    At higher redshifts, the correlation is only visible in the most overdense regions. Shaded regions indicate the standard error on the mean in each overdensity bin.}

    \label{fig:mass_vs_density}
\end{figure*}

To quantify the mass–density relation, we show the average stellar mass as a function of overdensity in Figure~\ref{fig:mass_vs_density}. The left panel shows results for the full galaxy population, the middle panel shows SFGs only, and the right panel shows QGs. In all cases, we observe a general trend where stellar mass increases with overdensity, although the trend is weaker in the SFG sample. The nature of this positive correlation between mass and density trend evolves with time. At lower redshifts ($z < 2.5$), the correlation is strong and evident even at moderate overdensities. Galaxies in overdense regions with $\log(1+\delta)\gtrsim 0.3$ tend to be about $0.3$–$0.6$~dex (a factor of $2$–$4$) more massive than those in average or underdense environments. This is consistent with expectations from hierarchical structure formation, where overdense environments host earlier and more rapid mass assembly. In the star-forming sample, the trend is still present but weaker, and the average mass is lower compared to QGs. This shows that massive galaxies in dense environments are more likely to be quiescent, which is a consequence of merging at later epochs.

At higher redshifts ($z>2.5$), the trend persists but becomes limited to the highest-density environments, and is dominated by SFGs. The stellar mass–density correlation is no longer present across all environments but only at the upper end of the overdensity distribution (e.g., $\log(1+\delta)\gtrsim 0.4$), where the mass difference remains small ($\lesssim0.2$~dex). This suggests that only the most extreme peaks of the matter density field, such as proto-clusters, host galaxies that have grown significantly in mass by those epochs. The reduced statistical range of environments at early times may also contribute to the flattening of the trend in low and intermediate density regions. This implies that the underlying physical processes such as biased halo formation, gas accretion, and merging are in place from early times, but their influence is confined to dense regions in the early universe. As cosmic structures grow and evolve, these processes extend to more typical environments, producing a smoother and more extended mass–density correlation at later times.

Our mass-environment trends are consistent with other studies. In the CANDELS field, \citet{Chartab2020} found that more massive galaxies preferentially reside in overdense regions, but the strength of the correlation declines toward higher redshifts. They also found that the correlation is stronger with QGs and weaker for SFGs. With the COSMOS2020 data, \citet{Taamoli2025} reported a similar positive mass-density relation up to $z\sim4$ with a gradual weakening at earlier epochs. They also note that at $z>2$, where the peak of star formation activity takes place, only the most extreme overdensities host galaxies that have already assembled significant stellar mass, consistent with our observation that the correlation at early times is limited to the upper tail of the density distribution. Compared to other works, COSMOS-Web's depth provides better mass completeness at high redshift, allowing us to trace the mass-density relation into regimes (low mass, high $z$) that were inaccessible in previous surveys. This combination of completeness and more accurate environment detection strengthens the statistical significance of the trends observed for COSMOS-Web, particularly at higher redshifts.

The positive correlation between the stellar mass and environment at low redshift can be explained within the hierarchical structure formation framework, where the highest peaks in the density field collapse earlier, leading to accelerated mass assembly \citep{White1978, Mo1996}. In these regions, the higher frequency of galaxy mergers \citep{Fakhouri2010, Lotz2011, Giddings2025} contributes to the buildup of massive systems, particularly quiescent galaxies at later times. Once dark matter halos reach $\sim10^{12}\,M_\odot$, infalling gas is shock-heated and the accretion mode changes from cold to hot \citep{Dekel2006, Birnboim2003}, reducing the supply of fresh gas. Combined with feedback from active galactic nuclei \citep{Croton2006, Fabian2012}, this leads to quenching while preserving high stellar mass. A significant fraction of galaxies are also affected by pre-processing in group environments before entering clusters, where tidal interactions and mergers increase their stellar mass \citep{Wetzel2013}. At higher redshift, dense filaments feeding proto-clusters can sustain high cold gas accretion rates \citep{Keres2005, Dekel2009, Daddi2022}. This allows galaxies in the most overdense regions to assemble large stellar masses earlier than those in lower-density environments, consistent with our $z>2.5$ results.

The mass–density relation can also be influenced by star formation activity and the efficiency of mass and environment quenching processes. Since both SFR and quenching affect the rate at which galaxies build up stellar mass, differences in these processes with density can shape part of the observed trends. These connections are explored further in Sections ~\ref{sec:SFR_Density} and ~\ref{sec:quenching_roles}.

\subsection{Evolution of SFR and sSFR with Large-Scale Environment Density} \label{sec:SFR_Density}

\begin{figure*}
    \centering
    \includegraphics[width=1\linewidth]{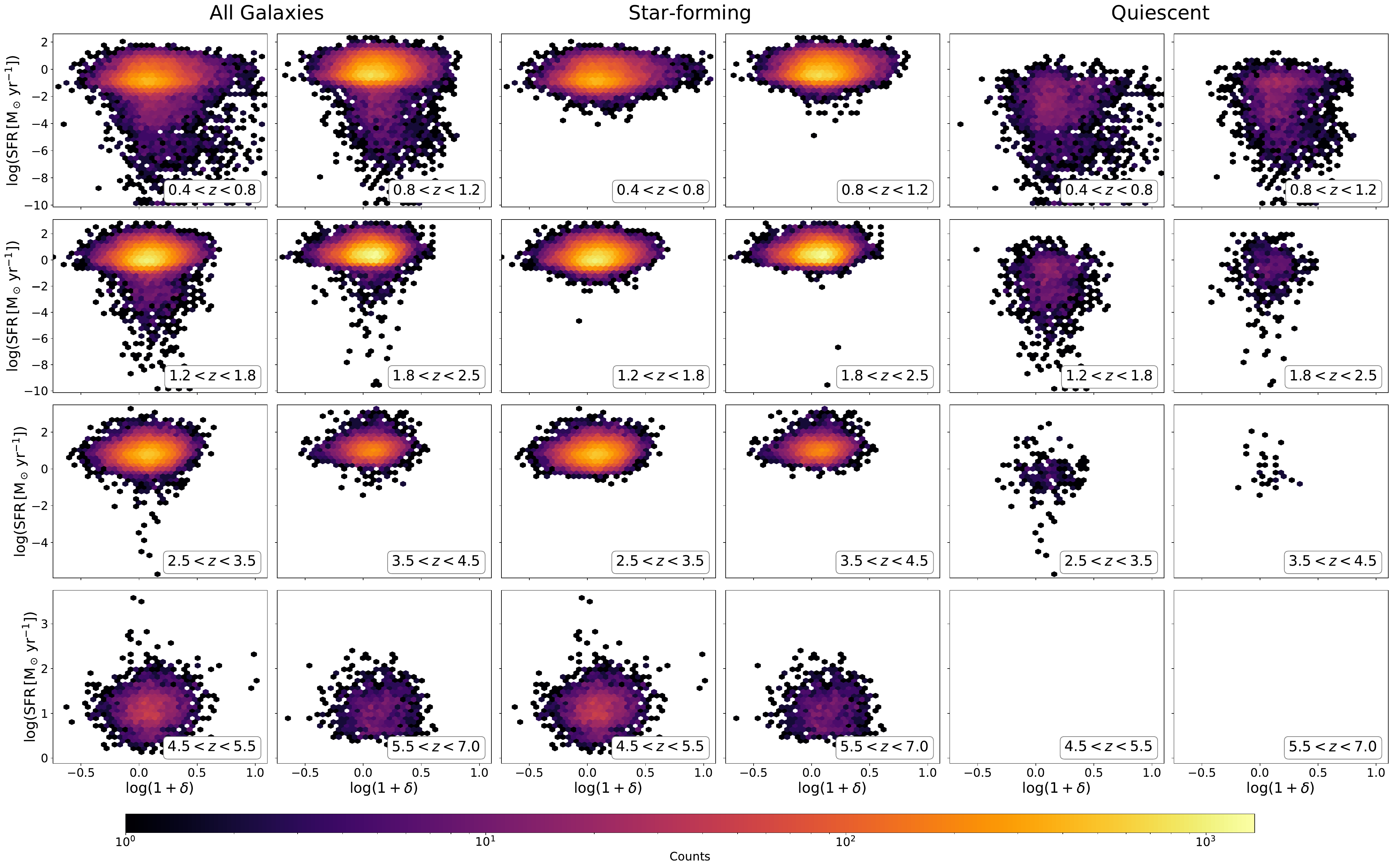}
    \caption{SFR as a function of environmental overdensity in COSMOS-Web. 
    Each row corresponds to a different redshift bin, increasing from top to bottom, and each column corresponds to a galaxy population: all galaxies (left), star-forming galaxies (middle), and quiescent galaxies (right). 
    Points are colored by number counts. 
    The distribution becomes narrower with increasing redshift, and the quiescent sample is sparse at $z \gtrsim 2.5$.}
    \label{fig:scatter_sfr_vs_density}
\end{figure*}

Studying how SFR and sSFR vary with large-scale structure density helps us understand whether galaxies form stars differently depending on their surrounding environment. Figure \ref{fig:scatter_sfr_vs_density} shows the distribution of SFR as a function of overdensity across redshift bins for our three samples. For the full sample at low redshift ($z < 1.8$), there is a clear tendency for galaxies with higher SFR to occupy lower-density regions, while high-density environments host fewer galaxies with strong star-formation rate. This trend is more pronounced at lower redshift and gradually weakens with increasing redshift. The star-forming sample exhibits a much flatter distribution across environments in the same redshift range, which suggest environments have a mild effect on the SFR of actively SFGs. Meanwhile, the quiescent population shows a negative correlation with density at $z<0.8$. This correlation diminishes at higher redshifts. Therefore, the trend seen in the full sample is primarily driven by the growing quiescent population in overdense regions at lower redshifts, which suppresses the average SFR in the total sample. 

\begin{figure*}
    \centering
    \includegraphics[width=1\linewidth]{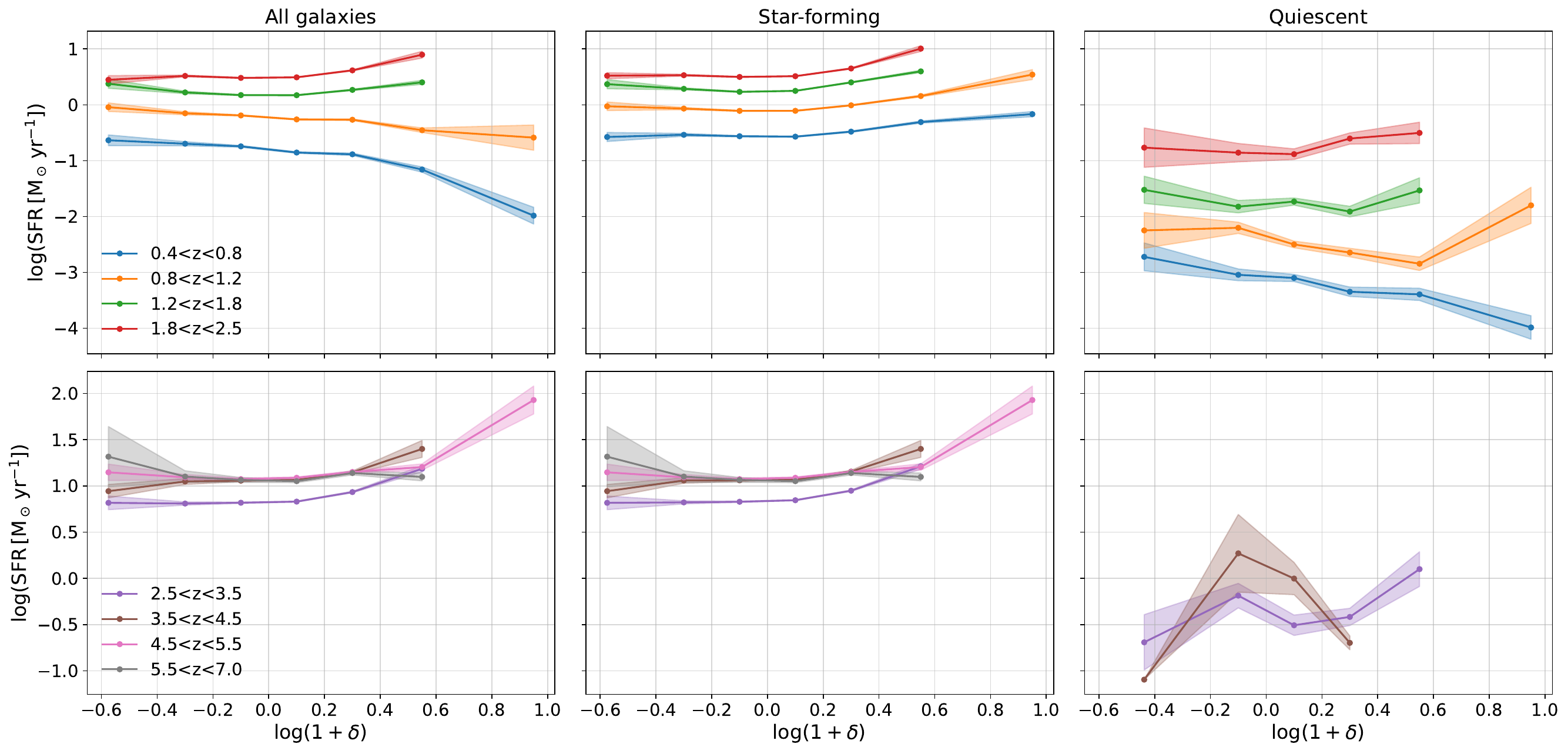}
    \caption{Average SFR as a function of environmental overdensity in COSMOS-Web. 
    Columns show all galaxies (left), star-forming galaxies (middle), and quiescent galaxies (right). 
    The top row corresponds to redshift bins in the range $0.4 < z < 2.5$, and the bottom row to $2.5 < z < 7.0$. 
    For the full population, SFR decreases with overdensity at $z \lesssim 1.2$, driven mainly by the higher quiescent fraction in dense regions. 
    SFGs instead show a mild increase of SFR with density up to $z \sim 5.5$, while QGs display a negative trend at $z \lesssim 1.2$, and a mild increase at highest densities at $z > 1.2$. Standard uncertainties are indicated with shaded regions.}
    
    \label{fig:sfr_vs_density}
\end{figure*}

Figure \ref{fig:sfr_vs_density} shows the average trend of SFR with LSS density. For the full sample (left panels), the trend is negative at $z < 1.2$, with SFR decreasing steadily toward higher densities. In the $1.2 < z < 1.8$ bin, this decline flattens, and by $z > 1.8$ the trend reverses and galaxies in overdense environments tend to have higher SFRs than those in underdense regions. Breaking the sample into galaxy types reveals the cause of the evolution in the full sample trend. In the star-forming sample (middle panels), SFR shows a mild positive correlation with density up to $z < 5.5$, with galaxies in denser regions forming stars at rates up to 0.5 dex higher, and the trend diminishes in the early epoch.  In contrast, the quiescent population (right panels) displays a negative environmental dependence for $z<0.8$. In the $0.8<z<1.2$ epoch, SFR decreases about 1 dex with density up to $\log(1+\delta)<0.6$, but SFR increases again at the highest density regions. This upturn in SFR is likely due to more gas reservoir in densest region. Also, these dense environments may host recently quenched galaxies with residual star formation, or  galaxies with brief rejuvenation due to minor mergers or tidal interactions \citep{Wild2016,Rutkowski2025}. Moreover, there is no clear evolution for QGs at higher redshifts. So, the decline in SFR with density for the full sample is just coming from QGs having lower SFR values in general.

\begin{figure*}
    \centering
    \includegraphics[width=1\linewidth]{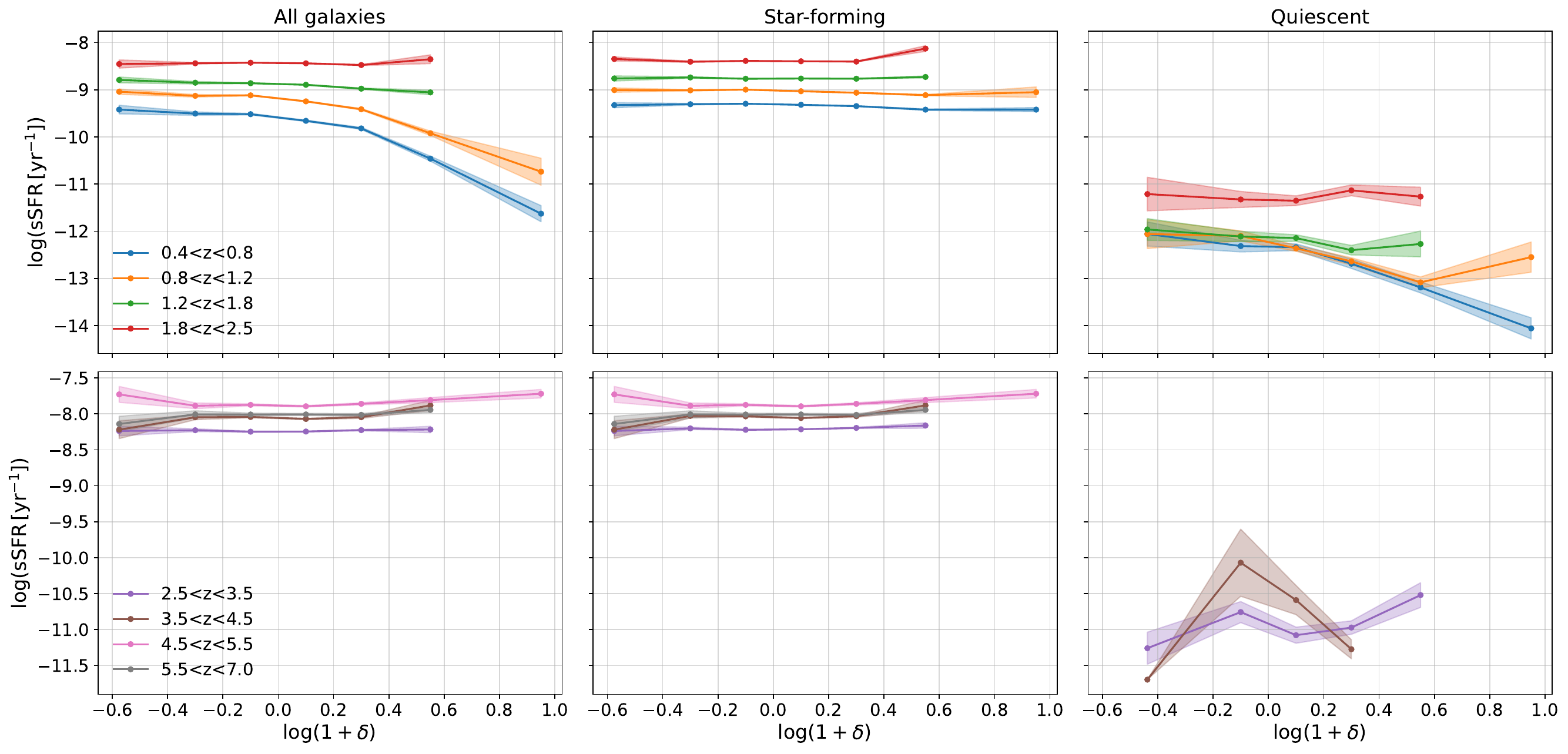}
    \caption{Average sSFR as a function of environmental overdensity in COSMOS-Web. 
    Columns show all galaxies (left), star-forming galaxies (middle), and quiescent galaxies (right). 
    The top row corresponds to redshift bins in the range $0.4 < z < 2.5$, and the bottom row to $2.5 < z < 5.0$. 
    At low redshift, sSFR declines in overdense regions due to QGs, while SFGs maintain nearly flat sSFR across environments. 
    At higher redshifts, the dependence on environment largely disappears for both populations. Standard uncertainties are shown with shaded areas.}
    \label{fig:ssfr_vs_density}
\end{figure*}

Figure~\ref{fig:ssfr_vs_density} shows how sSFR changes with overdensity for different redshift bins and galaxy types. For the star-forming sample, sSFR stays nearly constant across all environments and redshifts. This is expected since both the SFR and stellar mass of these galaxies increase with density, keeping the ratio roughly the same. Similar flat trends have been reported in other studies, showing that the environment does not have a significant effect on sSFR for SFGs \citep{Scoville2013, Darvish2016, Taamoli2024}.

At low redshifts ($z<1.2$), sSFR for QGs decreases with increasing density, which is consistent with a drop in SFR and an increase in mass. This trend supports the idea that environmental quenching, such as gas stripping or starvation, suppresses residual star formation in dense environments \citep{Peng2010, Woo2013}. In the $0.8 < z < 1.2$ bin, sSFR increases again at the highest densities. This may be due to recently quenched galaxies with low-level star formation or temporary rejuvenation events triggered by interactions or mergers \citep{Wild2016,Rutkowski2025}. Above $z > 1.2$, there is no clear trend, suggesting that the effect of environment on sSFR becomes weaker at early times \citep{Taamoli2024, Taamoli2025}. In the full sample, we see a negative trend for sSFR and overdensities up to $z<1.2$, but it is largely caused by evolution of QGs.

Our findings broadly agree with recent studies. At low redshift ($z \lesssim 1.2$), our negative correlation between SFR and overdensity is consistent with the results of \citet{Chartab2020} and \citet{Taamoli2024}, both of which report significant suppression of SFR in denser regions out to $z \sim 1.1$. The flattening we observe at $1.2 < z < 1.8$ matches the transitional regime identified by \citet{Taamoli2024}, where the correlation weakens and becomes statistically insignificant. Beyond $z \gtrsim 1.8$, we find a clear reversal in correlations. Galaxies in overdense regions exhibit enhanced SFR, which is in agreement with \citet{Lemaux2022} and \citet{Taamoli2024}, who both report a monotonic increase in SFR (and sSFR) with density at $z > 2$. However, the amplitude of our reversal appears somewhat smaller than in \citet{Lemaux2022}, who find up to an order-of-magnitude enhancement between the lowest and highest densities. This may reflect differences in sample selection or the range of probed environments. Compared to \citet{Chartab2020}, that report no statistically significant reversal up to $z \sim 3.5$, our detection likely benefits from deeper NIR data and improved photometric redshift precision in COSMOS-Web, and larger sky area coverage compared to CANDELS, that enabled more reliable density estimation at higher redshifts.

The negative correlation between SFR and LSS density at low redshifts ($z\lesssim1.2$) is consistent with environmentally driven quenching processes that are more effective in dense regions \citep{Taamoli2025}. We will discuss how quenching efficiencies can replicate SFR trends with density in Section \ref{sec:quenching_roles}. Mechanisms such as ram pressure stripping can rapidly remove the cold gas reservoir from galaxies moving through cluster environments  \citep{Gunn1972, Boselli2006}. This process can shut down star formation on short timescales ($\sim 1\text{Gyr}$). Strangulation or starvation \citep{Moore1999, Peng2015} suppresses star formation more gradually by cutting off the supply of fresh cold gas from the cosmic web, leading to quenching over $\sim 1-3\text{Gyr}$. Moreover, tidal harassment from repeated high-speed encounters can both strip gas and dynamically heat the stellar disk, which indirectly reduces star formation efficiency \citep{Moore1996}. In addition, AGN feedback, which is more prevalent in massive galaxies in dense environments, can heat or expel the interstellar medium, thereby suppressing star formation and reinforcing quenching in these regions \citep{Fabian2012, Terrazas2020}.

At higher redshifts ($z \gtrsim 1.8$), the reversal of the SFR-density relation suggests that dense environments may instead enhance star formation activity. One of the contributing factors for this reversal is that SFGs dominate the sample at higher redshifts. Another key factor behind this is that galaxies at early times contained a much larger reservoir of cold gas.  \citet{Tacconi2010} found that the cold gas made up roughly $44\%$ of the baryonic mass in galaxies at $z \sim 2.3$, and about $34\%$ at $z \sim 1.2$, which is higher than in the local Universe. These large gas supplies provided abundant fuel for star formation. Another is the supply of cold gas from large-scale structure filaments feeding proto-clusters, which can sustain high SFRs \citep{Dekel2009}. Dense environments at these epochs also tend to have higher rates of galaxy–galaxy interactions and mergers, which can sometimes trigger starbursts \citep{Snyder2017, Ventou2019, Giddings2025}. Moreover, the rapidly evolving gravitational potential in assembling structures can tidally compress gas, further inducing bursts of star formation \citep{Bekki1999}. Also, the cosmic web is not matured enough to quench galaxies as efficient as lower redshifts \citep{Taamoli2025}. Quiescent galaxies are responsible for the SFR-density trend in general at low redshifts. Therefore, with galaxies being mostly starburst at higher redshifts, the SFR-density trend becomes dominated by SFGs, which have a positive correlation with density.

The nearly flat sSFR–density relation for SFGs across most redshifts implies that the environmental dependence of their star formation rates is closely coupled to their stellar mass growth. In denser regions, both SFR and $M_\ast$ increase in parallel, keeping sSFR roughly constant. This is consistent with scenarios where the same processes that enhance gas supply (e.g., cold gas accretion along filaments, merger-driven inflows) also drive faster stellar mass assembly \citep{Dekel2009, Genzel2015, Speagle2014}. For QGs, the negative sSFR–density trend at low $z$ reflects both suppressed SFR and higher masses in denser regions, consistent with environmental quenching acting more strongly where galaxy interactions are prevalent \citep{Taamoli2025, Chartab2020}. Since sSFR measures the relative growth rate of stellar mass, its environmental dependence directly links the SFR trends discussed here to the mass–density relation in Section~\ref{sec:mass_density}. These variations in sSFR indicate differences in how efficiently QGs and SFGs build mass in different environments, helping to explain part of the mass–environment correlation. We explore this relation in Section \ref{sec:quenching_roles} to quantify the effect of mass and environment.

Table \ref{tab:corr_binned} presents Pearson correlation coefficients between $\log(1+\delta)$ and stellar mass, SFR, and sSFR. In each redshift interval, galaxies are binned by $\log(1+\delta)$ using the same density bins as in Figs.~\ref{fig:mass_vs_density}, \ref{fig:sfr_vs_density}, and \ref{fig:ssfr_vs_density}. For each density bin and for each sample (All, SFG, QG) we compute the mean of the corresponding quantity, and we then evaluate the correlation between those binned means and $\log(1+\delta)$. The table lists the coefficients together with the sample size $N$ and the 10th percentile stellar mass $\log(M_{10\%}/M_\odot)$. The results are consistent with the trends shown in Figs ~\ref{fig:mass_vs_density}, \ref{fig:sfr_vs_density}, and \ref{fig:ssfr_vs_density}. The highest redshift bin ($5.5<z<7.0$) has a small sample size, and the scatter plots of physical parameters (e.g., stellar mass, SFR, sSFR) do not show strong correlations with density. Therefore, the correlation coefficients should be interpreted with caution, even though the p-values are small ($P<0.01$) for this bin.

\begin{table*}[t]
\centering
\footnotesize
\setlength{\tabcolsep}{3pt}%
\begin{tabular*}{\textwidth}{@{\extracolsep{\fill}} @{} l c c ccc ccc ccc @{} }
\toprule
 & & & \multicolumn{3}{c}{Correlation Mass vs Density} & \multicolumn{3}{c}{Correlation SFR vs Density} & \multicolumn{3}{c}{Correlation sSFR vs Density} \\
\cmidrule(lr){4-6} \cmidrule(lr){7-9} \cmidrule(lr){10-12}
Redshift Range & $N$ & $\log(M_{10\%}/M_\odot)$ & All & SFG & QG & All & SFG & QG & All & SFG & QG \\
\midrule
$0.4 < z < 0.8$ & 21535 & 8.11 & 0.89 & 0.89 & 0.81 & -0.90 & 0.90 & -0.98 & -0.90 & -0.84 & -0.96 \\
$0.8 < z < 1.2$ & 33075 & 8.22 & 0.86 & 0.85 & 0.90 & -0.98 & 0.80 & 0.13 & -0.92 & -0.71 & -0.72 \\
$1.2 < z < 1.8$ & 34254 & 8.29 & 0.65 & 0.58 & 0.49 & 0.12 & 0.58 & -0.15 & -0.97 & 0.31 & -0.86 \\
$1.8 < z < 2.5$ & 33990 & 8.24 & 0.84 & 0.87 & 0.81 & 0.82 & 0.76 & 0.70 & 0.51 & 0.55 & 0.07 \\
$2.5 < z < 3.5$ & 20163 & 8.37 & 0.82 & 0.80 & 0.33 & 0.81 & 0.82 & 0.74 & 0.48 & 0.85 & 0.74 \\
$3.5 < z < 4.5$ & 6473 & 8.49 & 0.71 & 0.69 & -0.13 & 0.90 & 0.90 & 0.32 & 0.86 & 0.85 & 0.27 \\
$4.5 < z < 5.5$ & 3243 & 8.39 & 0.79 & 0.79 & -- & 0.74 & 0.74 & -- & 0.26 & 0.26 & --  \\
$5.5 < z < 7.0$ & 1121 & 8.58 & -0.70 & -0.70 & -- & -0.58 & -0.58 & -- & 0.83 & 0.83 & -- \\
\bottomrule
\end{tabular*}
\caption{Correlation coefficients vs. density (binned by $\log(1+\delta)$)}
\label{tab:corr_binned}
\end{table*}

\subsection{The Roles of Environment and Mass in Galaxy Quenching} \label{sec:quenching_roles}
The trends in quiescent galaxy physical parameters with overdensity presented in the previous subsections suggest that both stellar mass and large-scale environment play a role in shutting down star formation. To examine how galaxies are distributed across different environments, we divide the density field into three regimes based on their overdensity. Galaxies are sorted by their density contrast, and the $P_{25}$ and $P_{75}$ percentiles are computed, and we find low, medium, and high-density regions in each redshift bin.
\begin{itemize}
\item \textbf{Low density:} $\delta \le P_{25}$
\item \textbf{Medium density:} $P_{25} < \delta \le P_{75}$
\item \textbf{High density:} $\delta > P_{75}$
\end{itemize}

\begin{figure}
    \centering
    \includegraphics[width=0.9\linewidth]{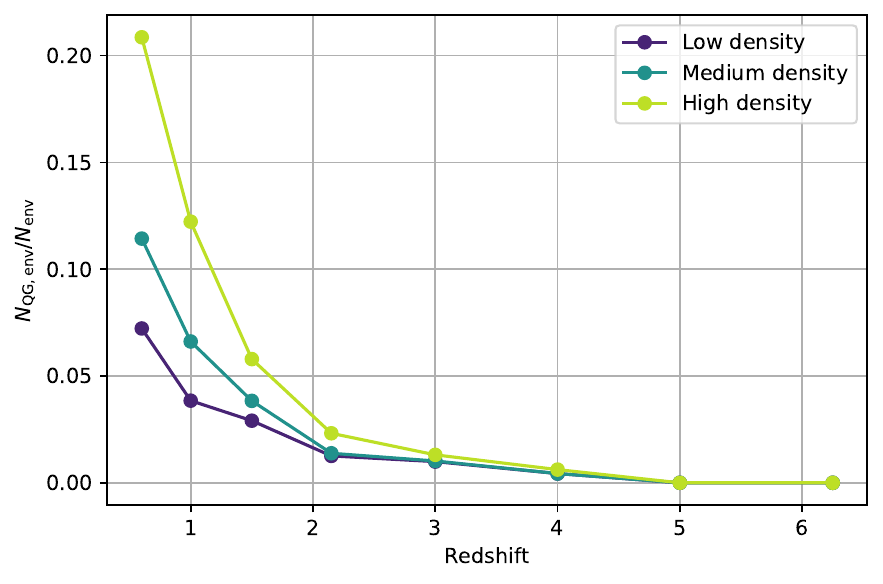}
    \caption{Quiescent fraction as a function of redshift in different environments. Curves show $f_{\mathrm{QG,env}} = N_{\mathrm{QG,env}}/N_{\mathrm{env}}$ measured within each redshift bin for
    low-density ($\delta \leq P_{25}$), medium-density ($P_{25} < \delta \leq P_{75}$), and high-density ($\delta \geq P_{75}$) regions. $N_{\mathrm{QG,env}}$ is the number of QGs in a given environment, and $N_{\mathrm{env}}$ is the number of galaxies in that environment. Density percentiles ($P_{25}$, $P_{75}$) are computed separately in each redshift bin from the overdensity distribution. The fraction of QGs in dense environments is higher at all redshifts.}
    \label{fig:quiescent_fraction_env_percentile}
\end{figure}

As shown in Figure \ref{fig:quiescent_fraction_env_percentile}, quiescent fraction increases with time and at later epochs, high-density regions have a higher quiescent fraction compared to other environments. However, to disentangle these effects, we follow the framework introduced by \citet{Peng2010}, where galaxy quenching is described as the combination of two independent processes, mass quenching and environmental quenching.  In this picture, the probability that a galaxy is quenched can be factorized into contributions from each process \citep{Peng2010, Peng2012, Darvish2016, Kawinwanichakij2017, Chartab2020, Lemaux2022, Taamoli2025}.

Mass quenching is generally associated with internal processes that scale with stellar mass or halo potential, such as virial shock heating in massive halos \citep{Dekel2006}, feedback from active galactic nuclei \citep{Croton2006, Fabian2012}, morphological quenching \citep{Martig2009}, or the exhaustion of cold gas reservoirs. Environmental quenching is instead linked to processes induced by the surrounding large-scale structure, including ram-pressure stripping \citep{Gunn1972}, harassment and tidal interactions, strangulation, and suppression of cold gas accretion in dense regions \citep{Moore1996, Dekel2009}.

To quantify these processes, we measure the quenching efficiencies following \citet{Peng2010, Chartab2020, Taamoli2025}. The environmental quenching efficiency is defined as:
\begin{equation}
\epsilon_{\mathrm{env}}(M_\star,z) = 1 - \frac{f_{\mathrm{SF}}(\delta \ge P_{75},\,M_\star,\,z)}{f_{\mathrm{SF}}(\delta \le P_{25},\,M_\star,\,z)},
\label{eq:eps_env_code}
\end{equation}
where $f_{\mathrm{SF}}=1-f_{\mathrm{QG}}$ is the star-forming fraction. This measures the fraction of galaxies that would be star-forming in low-density regions but are quenched in denser environments.

Similarly, the mass quenching efficiency is defined as:
\begin{equation}
\epsilon_{\mathrm{mass}}(M_\star,z \mid \delta\ge P_{75}) = 1 - \frac{f_{\mathrm{SF}}(M_\star,\,\delta \ge P_{75},\,z)}{f_{\mathrm{SF}}(M_0,\,\delta \ge P_{75},\,z)},
\label{eq:eps_mass_code}
\end{equation}
where $M_0$ is the lowest stellar mass bin in the same redshift slice. This measures the fraction of galaxies that would be star-forming at low mass but are quenched at higher mass, at fixed environment. We also define a normalized difference between the two effects to see which one is more dominant as follow:
\begin{equation}
\Delta_{\mathrm{norm}}(M_\star,z) = \frac{\epsilon_{\mathrm{env}}-\epsilon_{\mathrm{mass}}}{\epsilon_{\mathrm{env}}+\epsilon_{\mathrm{mass}}},
\end{equation}

\begin{figure*}
    \centering
    \includegraphics[width=1\linewidth]{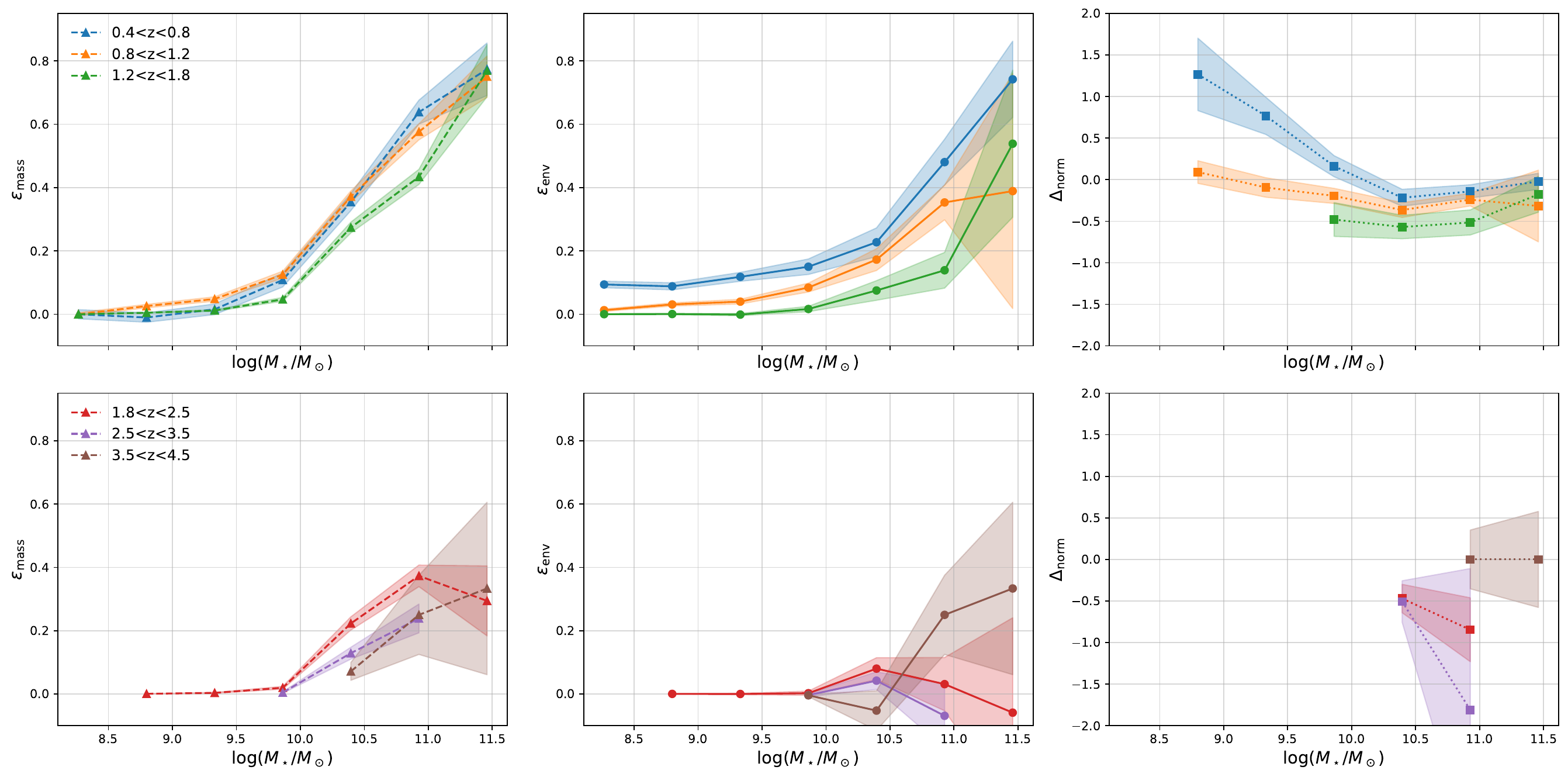}
    \caption{Quenching efficiencies as a function of stellar mass in different redshift bins. The top panels are showing redshift bins in the range $0.4<z<1.8$, and bottom panels are showing $1.8<z<4.5$.
    The left panels show the environmental quenching efficiency, comparing the fraction of quenched galaxies in high- versus low-density environments. 
    The middle panels present the mass quenching efficiency, showing how quenching depends on stellar mass within dense regions. 
    The right panels show the normalized strength, which quantifies whether quenching is primarily driven by environment (positive values) or by stellar mass (negative values). 
    Shaded regions represent uncertainties, which are binomial errors. For the normalized difference, errors are propagated from mass and environment efficiencies, and only bins with relative uncertainty $(\partial\Delta_{\mathrm{norm}}/\Delta_{\mathrm{norm}}) < 3$ are shown. Mass quenching dominates at most redshifts and masses, while environmental quenching becomes stronger at $z < 0.8$ for low-mass galaxies.}
    \label{fig:quenching_efficiencies_vs_mass_normdiff}
\end{figure*}

The quenching efficiencies and the normalized difference between the two effects are presented in Figure~\ref{fig:quenching_efficiencies_vs_mass_normdiff}. The earliest redshift bin showing non-zero values for both $\epsilon_{\mathrm{mass}}$ and $\epsilon_{\mathrm{env}}$ is $3.5<z<4.5$. However, due to the limited number of QGs at this epoch, it is not possible to determine which channel is dominant. At early times, quenching appears first at higher masses, and as galaxies evolve, at later times, quenched galaxies become more common at lower masses as well. This is consistent with the downsizing scenario where galaxy quenching proceeds from high- to low-mass systems as the universe evolves \citep{Cowie1996, Fontanot2009}. This trend may be partly influenced by sample limitations and selection effects, since detecting low-mass quiescent galaxies at high redshift is challenging. At all redshifts, $\epsilon_{\mathrm{mass}}$ increases with stellar mass, as expected if more massive galaxies are more strongly affected by mass-dependent quenching mechanisms. The first epoch in which mass quenching shows a dominant effect is $2.5<z<3.5$, where the mass efficiency is exceeding environmental quenching effects. Environmental quenching efficiencies also rise with stellar mass, but generally with smaller amplitude. Comparing different epochs, both $\epsilon_{\mathrm{mass}}$ and $\epsilon_{\mathrm{env}}$ increase toward lower redshift, reflecting the growing impact of quenching over cosmic time. At high redshift, quenching is dominated by the mass channel, with environment playing only a minor role. For the normalized efficiencies, we show only bins with relative uncertainty ($\partial\Delta_{norm}/\Delta_{norm}$) less than $3$. Above $z\sim1.8$, mass quenching dominates across the full stellar mass range. Even low-mass galaxies in this regime are more strongly affected by mass-related processes than by environmental factors. In the $0.8<z<1.2$ bin, the environmental quenching strength becomes comparable to mass quenching efficiency for low-mass galaxies, but mass quenching becomes dominant at higher masses. Below $z\sim0.8$, however, the balance shifts. Environmental quenching becomes more effective than mass quenching at the low-mass end ($M_\star<10^{10}M_\odot$), indicating that dense environments are increasingly capable of shutting down star formation in relatively small systems through mechanisms such as ram-pressure stripping or strangulation. With increasing stellar mass in this low-redshift epoch, the efficiency of environmental quenching also rises, eventually reaching values comparable to mass quenching at the highest masses. By the present epoch, the most massive galaxies in dense regions appear to be shaped by a combination of both processes operating at similar strengths.

The balance between mass- and environment-driven quenching evolves with stellar mass and redshift, marking a transition in galaxy evolution pathways. At earlier times, internal processes tied to mass are the primary driver of star formation suppression in galaxy populations, even though mass was not fully assembled. However, in the later times, the environment exerts a stronger influence, making environmental quenching more dominant for low-mass systems and eventually matching mass quenching in the most massive galaxies in the densest regions.

Compared to previous work, our COSMOS-Web measurements push the study of quenching to higher redshifts and lower masses. For instance, COSMOS2020 sample was very limited by its mass completeness and for the redshift bin of $1.1<z<2$, COSMOS2020 is missing masses lower than $10^{9.5}M_\odot$, and for $2<z<4$ masses lower than $10^{10}M_\odot$ are missing \citep{Taamoli2025}. This is also the case for the CANDELS field. Although the samples are not exactly comparable, the final outcome is mostly consistent. \citet{Taamoli2025} reported that for COSMOS2020 at $2<z<4$, mass quenching dominates while environmental quenching efficiencies become negative, which means overdense regions enhance star formation due to abundant cold gas and frequent mergers. In contrast, we already measure nonzero values for both quenching efficiencies at $z\sim4$, although limited statistics at these redshifts prevent a robust ranking between them. This suggests that environmental effects may set in earlier than inferred from COSMOS2020, but both studies agree that mass quenching remains the stronger process in general at higher redshifts. Similar conclusions were found by \citet{Chartab2020}, who found that in the CANDELS field environmental quenching is negligible above $z \sim 1.5$ and mass quenching dominates across all stellar masses. Moving into intermediate redshifts ($1 \lesssim z \lesssim 2$), all studies converge on the continued dominance of mass quenching, with environmental effects still weak or marginal \citep{Taamoli2025,Chartab2020}. In the ZFOURGE survey performed to $z\sim2$, \citet{Kawinwanichakij2017} found that environmental and mass quenching strengths are comparable for $M_\star>10^{10.5}M_\odot$ at all redshifts. At lower redshifts ($z \lesssim 1$), however, we observe a rise in environmental quenching efficiency at the low-mass end, which dominates the mass quenching processes even  for galaxies with $M_\star \lesssim 10^{10}M_\odot$. This is in line with the low-redshift behavior reported by \citet{Peng2010, Kawinwanichakij2017,Chartab2020,Taamoli2025}, who all found that environmental quenching becomes increasingly important at late times, particularly for low-mass systems likely to be satellites in dense environments \citep{Peng2012}. Also, \citet{Toni2025} found that the quiescent galaxy fraction increases strongly with group richness below $z\sim1$, highlighting the growing importance of dense environments in shutting down star formation at late times.

The observed interplay between mass, star formation, environment, and quenching shows a tightly connected, multi-dimensional framework of galaxy evolution. The mass–density relation (Fig \ref{fig:mass_vs_density}) shows that, over time, galaxies in denser environments tend to build up more stellar mass, particularly in quiescent systems. At low redshifts, the suppression of star formation in dense environments leads to a higher abundance of quenched galaxies. In contrast, at higher redshifts, galaxies in overdense regions exhibit elevated star formation rates, likely fueled by abundant cold gas. These environmental effects are reflected in the quenching efficiencies. At high redshift, mass quenching dominates across all environments, consistent with internal mechanisms regulating star formation as galaxies grow in mass. Toward lower redshifts, environmental quenching rises in importance, especially for low-mass systems. This shows the impact of dense surroundings on gas removal and accretion suppression. Taken together, our results suggest that the pathways by which galaxies grow and shut down star formation are fundamentally shaped by a combination of intrinsic properties (mass, gas content) and extrinsic conditions (environment), with their relative importance evolving over cosmic time. This integrated view connects the physical processes driving the trends seen in stellar mass buildup, star formation activity, and quenching, offering a comprehensive picture of how galaxies transition from star forming to quiescent and evolve across environments and epochs.

\section{Summary} \label{sec:Summary}
We reconstruct the cosmic web density field in the COSMOS-Web field up to redshift of $9.5$. Our selection criteria is summarized in Table \ref{table:sample_selection}. We choose a fixed comoving width of $35\,h^{-1}\,\mathrm{Mpc}$ for our slices, which gives 157 slices in our sample. By assigning weights to each galaxy for each slice based on the redshift PDF, we generate density maps for each slice using the wKDE method, taking into account edge effects, adaptive bandwidths, and corrections for star-masked regions. This approach allowed us to study the evolution of galaxies within different environments up to $z \sim 7$.

Compared to COSMOS2020, COSMOS-Web provides better mass completeness and improved, deeper photometric redshift precision. These advances lead to cleaner and more reliable reconstructions of large-scale structure, with overdense regions no longer artificially inflated, underdense regions better preserved, and small-scale features more accurately recovered. COSMOS-Web also includes a larger number of faint, high-redshift galaxies, which improves environmental statistics. While COSMOS2020 density maps were limited to about $z \sim 4$, COSMOS-Web allows us to extend the reconstruction reliably up to $z \sim 7$, offering a more accurate and deeper view of galaxy properties in environments.

We explore the relation between mass, SFR, sSFR, and quenching efficiencies with LSS densities at different epochs to see how environment shape the characteristics of galaxies. Our results are summarized as follows:

\begin{enumerate}
    \item The relation between stellar mass and environment shows that massive galaxies preferentially occupy overdense regions. This correlation is strongest at $z \lesssim 2.5$, particularly for quiescent galaxies, while star-forming systems show a weaker dependence. At higher redshifts, the trend persists only in the most extreme overdensities, suggesting that the densest regions (e.g., proto-cluster environments) are the main sites of early mass assembly.
    
    \item For the full population, SFR decreases with overdensity at $z \lesssim 1.2$, flattens at $1.2 < z < 1.8$, and reverses beyond $z \gtrsim 1.8$, where galaxies in dense regions show enhanced star formation. This overall trend is driven by quiescent galaxies, which suppress the mean SFR in dense regions at low redshift ($z<1.2$), while star-forming galaxies instead show a weak positive correlation with density up to $z \sim 5.5$. sSFR stays nearly flat for star-forming systems across all redshifts, but declines with density for quiescent galaxies at $z \lesssim 1.2$ as SFR decreases and mass increases, consistent with environmental quenching. At higher redshift, the environmental dependence of both SFR and sSFR largely weakens, though overdense regions continue to host the most actively star forming galaxies.

    \item Quenching efficiencies measured with COSMOS-Web show that both stellar mass and environment contribute to shutting down star formation, but their relative importance evolves with redshift. At high redshift ($z \gtrsim 2.5$), quenching is dominated by mass-related processes, with environment playing only a minor role. As redshift decreases, both efficiencies increase, and by $z < 0.8$ environmental quenching becomes stronger than mass quenching for low-mass galaxies ($M_\star \lesssim 10^{10}M_\odot$). In dense regions at low redshift, massive galaxies are governed by a combination of mass and environmental processes acting with comparable strength. These results indicate that internal mechanisms tied to mass dominate early quenching, while environmental effects grow at later times, especially for low-mass systems.
\end{enumerate}

The observed correlations between stellar mass, star formation activity, quenching processes, and LSS density indicate that galaxy evolution has been shaped by both intrinsic and extrinsic factors across cosmic time. These observed links between galaxy properties and environment are made more reliable by the depth and precision of COSMOS-Web, which sets a new benchmark for large-scale structure studies.

\section*{Data Availability}
The code developed for this work to construct the wKDE density maps is openly available on \textit{GitHub}: \href{https://github.com/hhatam/CosmicWeb}{hhatam/CosmicWeb}. The repository includes all scripts used to generate the density maps, density contrast of galaxies, and plotting each step of the method for redshift slices. The resulting catalog containing density maps and density contrast of galaxies are available for public. In addition, the repository contains a video that illustrates the evolution of large-scale structures across cosmic time. To further support outreach and interactive exploration, we have also released an augmented-reality tool that enables users to visualize the COSMOS-Web density field in three dimensions. This tool is designed as an educational resource, allowing users to inspect how galaxies populate overdense and underdense regions in the cosmic web.

\begin{acknowledgments}
This project has received funding from the European Union’s Horizon 2020 research and innovation programme under the Marie Skłodowska-Curie grant agreement No 101148925. J.R.W. acknowledges that support for this work was provided by The Brinson Foundation through a Brinson Prize Fellowship grant.
\end{acknowledgments}

\bibliography{cosmos}{}
\bibliographystyle{aasjournalv7}

\clearpage
\suppressAffiliationsfalse   % <— re-enable affiliations
\allauthors

\end{document}